\documentclass[twocolumn,aps,prx,amssymb,raggedbottom,nobalancelastpage,superscriptaddress]{revtex4}

\usepackage{amsmath}
\usepackage{amssymb}
\usepackage{amsfonts}
\usepackage{dsfont}
\usepackage{graphicx}
\usepackage{bm}
\usepackage{color}
\usepackage{appendix}
\usepackage{epsfig}

\newcommand\be{\begin{equation}}
\newcommand\ee{\end{equation}}
\newcommand\bea{\begin{eqnarray}}
\newcommand\eea{\end{eqnarray}}

\begin{document}
\title{Hidden order and flux attachment in symmetry protected topological phases: a Laughlin-like approach}

\author{Zohar Ringel}
\affiliation{Theoretical Physics, Oxford University, 1, Keble Road, Oxford OX1 3NP, United Kingdom.}
\author{Steven H. Simon}
\affiliation{Theoretical Physics, Oxford University, 1, Keble Road, Oxford OX1 3NP, United Kingdom.} %

\begin{abstract}
Topological phases of matter are distinct from conventional ones by their lack of a local order parameter. Still in the quantum Hall effect, hidden order parameters exist and constitute the basis for the celebrated composite-particle approach. Whether similar hidden orders exist in 2D and 3D symmetry protected topological phases (SPTs) is a largely open question. Here we introduce a new approach for generating SPT groundstates, based on a generalization of the Laughlin wavefunction. This approach gives a simple and unifying picture of some classes of SPTs in 1D and 2D, and reveals their hidden order and flux attachment structures. For the 1D case, we derive exact relations between the wavefunctions obtained in this manner and group cohomology wavefunctions, as well as matrix product state classification. For the 2D Ising SPT, strong analytical and numerical evidence is given to show that the wavefunction obtained indeed describes the desired SPT. The Ising SPT then appears as a state with quasi-long-range order in composite degrees of freedom consisting of Ising-symmetry charges attached to Ising-symmetry fluxes. 
\end{abstract}


\maketitle

\section{Introduction}
In the past three decades, topological phases of matter have attracted a large amount of interest due to their tendency to exhibit highly robust quantum phenomena which have various applications in quantum engineering and metrology. One of the current frontiers in the field aims at understanding the variety of novel topological phases which arise when some extra symmetries are not allowed to be broken. For example, considering the case of an Ising symmetry, one may ask whether new topological phases exist in spin systems, provided that no magnetic fields or magnetic impurities are allowed. For this case, it was shown~\cite{Chen2011,Levin2012} that there are two topologically distinct Ising paramagnets in both one and two dimensions (1D,2D). Both phases are ``integer" (or short range entangled) paramagnetic phases; however they are distinguished by the absence or presence of protected boundary excitations. By now a variety of integer and fractional (long range entangled) SPTs are known to be theoretically possible \cite{Chen2011,YuanMing2012,Kapustin2014,Schuch2011} along with some experimental realizations. The latter include 2D and 3D topological insulators \cite{Hasan2010}, the Haldane chain \cite{Buyers1986}, and recently an experiment realizing a 2D bosonic SPT has been proposed \cite{Wen2014}. 


At least in 2D, the quantum Hall effect (QHE) is the basic building block of most topological phases. For example, upon replacing the layer index by a spin index, a quantum Hall bilayer in which each layer experiences an opposite magnetic field (a $(1,\bar{1},0)$ state \cite{Halperin1983}) describes a topological insulator \cite{Hasan2010}. Similarly a modified-(0,0,1) quantum Hall bilayer state of bosons can be thought of as a bosonic SPT, protected by the $U(1)$ symmetry associated with the charge difference between the layers \cite{Senthil2013}. Furthermore, by condensing charges such that this symmetry is broken down to a discrete subgroup $G$, one can obtain a large variety of abelian 2D SPTs \cite{YuanMing2012,YuanMing2013}. Also in some recent works \cite{AshvinDecorated,Zaletel2014,Juven1,Juven2}, effects of flux and charge binding, analogous to those occurring in the QHE, have been established for several types of SPTs.   

Laughlin's wavefunctions \cite{Laughlin1983} are powerful tools for analyzing the fractional quantum  Hall effect. Besides being extremely good approximations for realistic Hamiltonians, they offer a simple picture of the groundstate and excitations of these phases, they embody the idea of flux attachment which is the basis of composite particle theory \cite{Zhang1992,Jain2007}, they allow the construction of parent Hamiltonians for which they are the exact ground state \cite{Haldane1983,Trugman1985}, they are amenable to numerical explorations \cite{Laughlin1983,Chakraborty1988,Zaletel2012}, and their connection with conformal field theories \cite{Moore1991} provides the basis for generalizing them to more exotic phases. 

The Laughlin wavefunctions also reveal an intriguing connection between topological phases and broken symmetry phases via the concept of hidden long range order. Indeed, with complex phases removed, Laughlin's wavefunctions appear as states in which the $U(1)$ charge symmetry is spontaneously broken \cite{Girvin1987,Read1989}. Curiously, a similar type of hidden long range order exists in some 1D SPTs\cite{Kennedy1992,Else2013,Quella2013}. The question of whether 2D SPTs show a hidden order structure, and if so of what kind, is a largely open one. 

Here we generalize the Laughlin wavefunctions to the SPT context by exchanging the $U(1)$ symmetry of these functions, by the symmetry, $G$, protecting the SPT. This results in Laughlin-like wavefunctions describing some bosonic SPTs in one and two dimensions. More specifically, we write an ansatz for the groundstate of SPTs on a lattice using a specific type of tensor product state (TPS). The TPS is written in the basis where $G$ acts diagonally (the symmetry-charge basis) and appears as a broken symmetry state dressed by a phase factor which attaches fluxes to charges of the symmetry $G$. Removing this latter phase factor the hidden order behind these functions is studied and identified. 

This work is organized as followed: We begin by considering the Laughlin wave function on a lattice and rewrite it as a TPS. This reformulation allows us to generalize the $U(1)$ symmetry of Laughlin's wavefunction, to a generic $Z_N$ symmetry. Doing so yields the TPS ansatz for the SPT groundstates which are the focus of this work.  Starting from 1D, for any $Z_N \times Z_N$ symmetry we use our ansatz to constructively obtain the groundstates of the $N$ possible SPTs in this class. The hidden order of these states is a simple spontaneous breaking of the discrete $Z_N\times Z_N$ symmetry. 
Generalized flux attachment unitary transformations which unveil this hidden order are also obtained. We further validate our construction either by presenting the states as matrix product states (MPSs) and appealing to the known classification \cite{Schuch2011} or by making a local unitary transformation which maps our states onto the ones obtained using group cohomology \cite{Chen2011}. It is also shown that these states obey a certain self-duality symmetry. 

In 2D we study the Ising ($Z_2$) SPT \cite{Levin2012,Chen2011b}, and again obtain a candidate state for the groundstate. Strong analytical and numerical evidence is given to show that it indeed describes the desired topological phase. Furthermore, in striking similarity with the Laughlin wavefunction \cite{Girvin1987,Read1989}, removing phases from this wavefunction in the Ising-charge basis ($\sigma_x$ basis, if the local Ising symmetry is given by $\sigma_x$) reveals a state with quasi-long-range order associated with a breaking of the Ising symmetry.  

\section{The TPS ansatz} 

To motivate our TPS ansatz, we begin by rewriting the Laughlin wavefunction for a finite size droplet as a TPS. Consider a bosonic Laughlin wavefunction at $1/m$ filling with $n$ particles  
\begin{align}
\Psi_{m}(z_1,...,z_n) &= \Pi_{i<j} (z_i - z_j)^{m} e^{ -\sum_i \frac{1}{4}|z_i|^2}. 
\end{align}
We introduce a $U(1)$ scalar field $\phi \in [0..1]$ and use it to decouple the particle-particle terms   
\begin{align}
\Psi_{m}(z_1,...,z_n) &= \int{\rm D} \phi e^{ -\int d^2 r \frac{\pi m}{2} (\nabla \phi)^2 + i  \pi m \phi (\rho(z)-\rho_0)} \Pi_i A_i \\ \nonumber 
A_i &= \delta \left(\int_{O_{z_i}} dl \nabla \phi - 1 \right), 
\end{align}
where $\rho_0$ is the neutralizing background charge, $\rho(z) = \sum_{i} \delta(z-z_i)$, $O_{z_i}$ is some infinitesimal circle around $z_i$, and we only allow vorticity around $z_i$'s and nowhere else. Similarly to how one may present the Laughlin wavefunction as a many-operator correlator in a chiral CFT \cite{Moore1991}, in the above we present it as a many-operator correlator in a non-chiral CFT (a $U(1)$ model) but choose operators which couple only to a chiral subset of observables. The above identity can be proven straightforwardly in the continuum limit using simple Gaussian integrations. To regularize ultraviolet divergences one may put it on a lattice. We refer the avid reader to App. (\ref{App:LaughlinTPS}) for its derivation.  

Next we restrict position to a dense lattice ($z_{ab}$, where $a$ and $b$ are the $x$ and $y$
coordinates), with spacing ($\epsilon$) much smaller than $1$ (the magnetic length), and discretize $\rho(z)$ and $\phi(z)$ to a lattice $\rho_{ab}$ taking values in the positive integers  (${\mathbb Z_+}$) and $\phi_{ab}\in [0,1)$. This yields
\begin{align}
\label{Eq:LaughlinTPS}
\langle \{ \rho_{ab} \} | \Psi_{m} \rangle &= \int \Pi_{ab} [d \phi_{ab}]  A_{\phi}(\rho_{ab}) \\ \nonumber 
A_{\phi}(\rho_{ab}) &= \delta \left(\int_{O_{z_{ab}}} dl \nabla \phi  - \rho_{ab} \right) e^{i \pi m \phi_{ab} (\rho_{ab}-\epsilon^2 \rho_0)} \\ \nonumber &\times e^{ -\frac{\pi m}{4} (\nabla \phi_{ab})^2} 
\end{align}  
where $|\{\rho_{ab}\}\rangle$ is a state in the occupation basis and derivatives should be understood as their lattice versions involving nearest neighboring lattice sites. Similarly $\int_{O_{z_{ab}}}$ denotes a discrete sum over a circle-like path of radius $r_0$, where $\epsilon \ll r_0 \ll 1$. The above expression now appears as a TPS with $\phi \in U(1)$ serving as the tensor index. An important and expected  \cite{Hastings2014} limitation of this TPS, is that the density of the lattice must scale with the droplet radius ($R$). Indeed since the circulation at radius $R$ would scale as $R^2$ (or particle number), the typical gradients at this radius would scale as $R$. Thus the lattice density would have to scale as $R^{-1}$ in order to prevent a breakdown of the continuum approximation. This mode of failure can be traced back to the chirality of the Laughlin state, manifest here in the fixed sign of all vortices. Notably in App. (\ref{App:FromSPT}) we show how arrive at a similar expression from a known wavefunction of a $U(1)$ SPT. Conveniently in this latter case there is no background charge and the net vorticity is zero. 

To carry Eq. (\ref{Eq:LaughlinTPS}) into the SPT context we note that the tensor indices are a $U(1)$ variable,  and that the charges ($\rho_{ab}$) generate ${\mathbb Z_+}$ upon addition, the latter being the positive subset of $U(1)$ representations ($Rep[U(1)] = {\mathbb Z}$). We thus replace the $U(1)$ symmetry by an Abelian discrete symmetry $G$, and replace charges by $Rep[G]$. Since $G$ representations have a cyclic group structure, there is no sense of restricting to positive representations. Accordingly, we allow all representations and remove the background charge ($\rho_0=0$). We also remove the $\phi$ field strength term ($(\nabla \phi)^2$), since as would be evident later, its main effect in the SPT context is to drive the TPS into a trivial state where charges are confined. Last, we determine the lattice derivative by demanding it to fulfill a consistency condition, given below, which is inspired by the notion of flux attachment. As shown below, this appears to work well both for 1D and 2D SPTs. 

We turn to write the TPS explicitly. We consider $Z_N$ groups and represent the group elements ($\phi$) as the numbers $0 \ldots N-1$ and group action as addition modulo $N$. A product $\phi_a \phi_b$ means adding $\phi_b$'s $\phi_a-$times.  Representations are labeled by the symmetry-charges $\alpha \in Z_N$ and for the character tables we use $\chi_{\phi}(\alpha) = e^{ \frac{2 \pi i}{N} \phi \alpha }$. Our Hilbert space is either a 1D lattice or a 2D triangular lattice where each site is occupied by a symmetry-charge degree of freedom, again thought of as an integer in $0 \ldots N-1$. Unlike some other approaches \cite{Chen2011}, we work in the on-site symmetry-charge basis $|\{ \alpha \}\rangle =  |\alpha_1\rangle | \alpha_2 \rangle ...$. On this basis the action of $g \in G$, on a site $i$ is diagonal and given by 
\begin{align}
\label{Eq:SymDef}
g|\alpha_i\rangle=\chi_{g}(\alpha_i)|\alpha_i\rangle.  
\end{align}
Tensor indices ($\phi_i$) are elements in $Z_N$ and also sit on sites rather than bonds. Our TPS ansatz for an SPT wavefunction in a phase $m$ ($|\psi_m \rangle$) is
\begin{align}
\label{Eq:TPS}
\langle \{ \alpha \} | \psi_{m} \rangle &= \sum_{\{ \phi\} \in G} \Pi_j A_{\phi_{\langle \cdot, j \rangle}}(\alpha_j) \\ \nonumber 
A_{\phi_{\langle \cdot, j \rangle}}(\alpha_j) &= \delta_{{\rm d}_d[\{ \phi \}]_j,\alpha_j} e^{i \theta_m(\phi_{\langle \cdot, j \rangle})}
\end{align} 
where $\phi_{\langle \cdot, j \rangle}$ denotes the nearest neighbors of $\phi_j$. The lattice derivative operator for dimension $d$ (${\rm d}_d[ \{ \phi \}]_j$), which is some function of $\phi_{ \langle \cdot, j \rangle}$, and the local phase factors ($e^{i \theta_m(\phi_{\langle \cdot, j \rangle})}$), are determined next. 
 
We demand the following generalized flux attachment condition on the overall phase factor of the TPS ($\Theta_m = \sum_j \theta_m(\phi_{\langle \cdot, j \rangle})$) 
\begin{align}
\label{Eq:FluxAttachment}
\Delta_{j} e^{i \Theta_m(\{\phi\})} \equiv \frac{e^{i \Theta_m(\{ \phi\}+ {\rm 1}_j)}}{e^{i \Theta_m(\{ \phi \} )}} &= \chi_{m}({\rm d}_d[\{ \phi \}]_j), 
\end{align}
where $\{ \phi\}+ {\rm 1}_j$ means taking $\{ \phi \} = \{ ..., \phi_{j-1},\phi_j,\phi_{j+1},...\}$ to $\{...\phi_{j-1},\phi_j+1,\phi_{j+1},...\}$. Note that the same condition is obeyed by Eq. (\ref{Eq:LaughlinTPS}) in the bosonic case. One can think of this equation as a (discrete) derivative of $e^{i \Theta_m(\{\phi\})}$ along path in configuration space of $\phi$'s. From this perspective $\chi_{m}({\rm d}_d[\{ \phi \}]_j)$ takes the role of a (discrete) differential of a single valued function ($e^{i \Theta_m(\{\phi\})}$) and must therefore be exact. This implies the following consistency condition 
\begin{align}
\label{Eq:Consistency}
\frac{\chi_{m}({\rm d}_d[\{ \phi \} + {\rm 1}_{j'}]_j) \chi_{m}({\rm d}_d[\{ \phi \}]_{j'})}{\chi_{m}({\rm d}_d[\{ \phi \} + {\rm 1}_j]_{j'}) \chi_{m}({\rm d}_d[\{ \phi \}]_{j})} &= 1,
\end{align}  
where $j$ and $j'$ are two neighboring sites. One can also write down an onsite consistency condition, requiring $N$ consecutive applications of Eq. (\ref{Eq:FluxAttachment}) on the same site to yield an overall factor of $1$. Provided that ${\rm d}_d[..]_j$ is not a function of site $j$ itself, this will be automatically obeyed since $(\chi_{m}(\alpha))^N = \chi_{mN}(\alpha) = \chi_{0}(\alpha) = 1$.  

Notably we do not claim that any solution of the flux attachment consistency equation yields an SPT (clearly the trivial ${\rm d}_d = 0$ operator is always solution). What we shall see below is that certain ``natural'' choices of ${\rm d}_d$ lead to known SPTs. 

\section{One dimension}

We proceed by solving the above consistency condition for a 1D lattice with a $Z_N$ degree of freedom on each site. As one can verify, a generic solution is a staggered derivative
\begin{align}
\label{Eq:D1}
{\rm d_1}[ \{ \phi \}]_{2j} &= \phi_{2j + 1} - \phi_{2j - 1}, \\ \nonumber 
{\rm d_1}[\{ \phi \}]_{2j+1} &= \phi_{2j } - \phi_{2j + 2}.
\end{align}
This equation along with Eq. (\ref{Eq:FluxAttachment}) determine $e^{i \Theta_m}$ up to a global phase. Explicitly we find that 
\begin{align}
\label{Eq:ThetaM}
\theta_m(\phi_{\langle \cdot,j \rangle}) &= 2 \pi m (-1)^j \phi_j \phi_{j+1},
\end{align}
plus some constant. The resulting wavefunction for the 1D SPT ground state in class $m$ is given by 
\begin{align}
| \psi_m\rangle &= \sum_{ \{ \phi \} } e^{\frac{2 \pi m}{N} i  \sum_j  (-1)^j \phi_j \phi_{j+1}} | \{ \alpha(\phi) \}\rangle 
\end{align}
where $\{ \alpha(\phi) \}$  denotes the charge configuration $\alpha_i = {\rm d_1}[\{ \phi \}]_i$. 

The above state generalizes the Laughlin wavefunction in the sense that it is written in the symmetry-charge basis and obeys flux attachment (Eq. (\ref{Eq:FluxAttachment})). As we now show, it also includes hidden long range order structure. In the case of the Laughlin wave function, hidden order is revealed by removing the complex phase from the wavefunction \cite{Girvin1987} in the charge basis. The transformation which does so, is the flux attachment transformation. We thus perform a similar procedure on our state. 

First note that on closed boundary conditions, with an even number of sites, $|\psi_m\rangle$ contains all possible charge configuration up to two global constraints. These are that the charges on the even and odd sites add to zero separately. These constraints simply ensure that the state is invariant under the global $Z_N$ symmetry acting on all even or odd sites, which can be understood as a $Z_N\times Z_N$ symmetry. 

Next define the wavefunction, $| | \psi_m | \rangle$, obtained by removing all complex phases from $|\psi_m\rangle$ in the symmetry-charge basis 
\begin{align}
| | \psi_m | \rangle &= \frac{1}{N^{L}} \sum_{ \{ \phi \} } | \{ \alpha(\phi) \}\rangle = \frac{1}{N^{L-1}}  \sum_{ \{ \alpha \}} \delta_{\sum \alpha,0} | \{ \alpha \} \rangle, 
\end{align}
where the Kronecker delta ($\delta_{\sum \alpha,0}$) enforces both of the above global constraints on the $\{ \alpha \}$ charge configuration, and we introduced the normalization factor on a ring with $2L$ sites ($\frac{1}{N^{L}}$). 

To test for off diagonal long range order (ODLRO) in $| | \psi_m| \rangle$ consider $R_{i,j} = \langle |\psi_m| | a^-_j a^+_i | | \psi_m | \rangle$ where $a^+_i$ ($a^-_j$) raises (lowers) the charge at point $i$ by one (for example $a^+_i | \alpha_i \rangle = | \alpha_i +1 \rangle$). As can be easily verified, both these operators have zero average on a state which respects the $Z_N\times Z_N$ symmetry. Consequently a long range behavior of $R_{i,j}$ implies hidden order associated with a breaking of the $Z_N\times Z_N$ symmetry. 

To show that $R_{i,j}$ is long ranged, we note that the action of $a^-_j a^+_i$ on a charge configuration respects the global constraint, or equivalently commutes with the $Z_N\times Z_N$ symmetry. Consequently on a ring with $2L$ sites we have 
\begin{align}
R_{i,j} &=  \langle |\psi_m| | a^-_j a^+_i | | \psi_m | \rangle \\ \nonumber &= \frac{1}{N^{2L-2}} \sum_{\{ \alpha \},\{ \alpha \} } \delta_{\sum \alpha,0} \delta_{\sum \alpha',0} \langle \{ \alpha' \} | a^-_j a^+_i | \{ \alpha \} \rangle = 1
\end{align}
where in the last equality we have used the fact that $a^-_j a^+_i$ acting on any constraints-respecting $\{ \alpha \}$, gives another constraints-respecting charge configuration.  

It is also possible to construct a unitary ``flux-attachment" transformation, acting within the eigenvalue $1$ subspace of the global $Z_N\times Z_N$ symmetry, which maps $|\psi_m\rangle$ to $| |\psi_m| \rangle$.  Explicitly one writes $U = \sum_{\{ \alpha \}} |\{\alpha \} \rangle  \langle \psi_m | \{ \alpha \} \rangle \langle \{ \alpha \}|$. Being diagonal in the symmetry-charge basis, this transformation commutes with the symmetry operation and furthermore removes all phases from $|\psi_m\rangle$ as required. We comment that $U$ preserves the locality of symmetry respecting operators, and in particular maps the Hamiltonian of the SPT to a local Hamiltonian which breaks the symmetry spontaneously. The ground state degeneracy of the ferromagnet reflects the degeneracy caused by the boundary states in the SPT. Furthermore $U$ coincides with the disentanglers obtained in Refs. \cite{Else2013,Quella2013}. 

The above properties strongly suggests that $|\psi_m\rangle$'s describe ground states of SPTs with a $Z_N \times Z_N$ symmetry in 1D. The rest of this section is devoted to proving this as well as showing the existence of a certain self-duality symmetry of $|\psi_m\rangle$. To this end, consider reversing the role of $\alpha$ and $\phi$ in the above TPS, such that $\phi$'s become the physical degrees of freedom and $\alpha$'s are traced over. By grouping elements in pairs, the resulting state $|\tilde{\psi}_m\rangle$ can be written as  
\begin{align}
\label{Eq:Wen1DState}
\langle \{ \phi \} | \tilde{\psi}_m\rangle = \Pi_{k} \nu_m(1,[\phi_{2k-1},\phi_{2k}],&[\phi_{2k+1},\phi_{2k+2}]) \\ \nonumber 
 \nu_m(1,[\phi_{2k-1},\phi_{2k}],[\phi_{2k+1},\phi_{2k+2}]) &= e^{\frac{2 \pi i m}{N} \phi_{2k} (\phi_{2k+1}-\phi_{2k-1})} 
\end{align}
Consider $[\sigma,\tau]$ as a group element in $Z_N \times Z_N$ and extend $\nu_m$ to be a function of all three arguments via symmetry ($\nu_m([\sigma_0,\tau_0],[\sigma_1,\tau_1],[\sigma_2,\tau_2]) = \nu_m(1,[\sigma_1-\sigma_0,\tau_1-\tau_0],[\sigma_2-\sigma_0,\tau_2-\tau_0]$). One then finds that $\nu_m(a,b,c)\cdot \nu_m(a,c,d) \cdot \nu^{-1}_m(b,c,d) \cdot \nu^{-1}_m(a,b,d) = 1$ for any $a,b,c,d \in Z_N \times Z_N$, and so $\nu_m$ are cocycles \cite{Chen2011} in the second cohomology group ${\rm H}^2(Z_N \times Z_N,U(1))$. Furthermore in App. (\ref{App:All}) we show that varying $m$, all cocycles of ${\rm H}^2(Z_N \times Z_N,U(1))$ are obtained. Thus $|\tilde{\psi_m}\rangle$ are just the ground states obtained in Ref. (\onlinecite{Chen2011}) for 1D SPTs with $G=Z_N \times Z_N$ with some specific choice of coboundary. Moreover they are related to our TPS via 
\begin{align}
\label{Eq:dRelation}
{\rm \hat{d}_1} | \tilde{\psi_m}\rangle &= | \psi_m\rangle, \\ \nonumber 
{\rm \hat{d}_1} &= \sum_{\{ \phi \}} | \{ {\rm d_1} [\{\phi\}] \} \rangle \langle \{ \phi \} |.  
\end{align} 
This algebraic relation is however short of implying topological equivalence between $|\psi_m\rangle$ and $|\tilde{\psi}_m\rangle$, since ${\rm \hat{d}_1}$ cannot be written as a product of local unitary transformations even for fixed boundary conditions (to see this, assume that it can be written in such a form. If so its inverse, ${\rm \hat{d}}^{-1}_1$, must also have a local unitary form whereas actually it is a highly non-local string-like transformation). 

Next we regroup the tensor and indices such that they appear as MPSs. This will allow us to use the known classification of MPSs \cite{Schuch2011}. Accordingly, the $\phi$'s and $\alpha$'s are paired into $[\sigma_k,\tau_k] = [\phi_{2k-1},\phi_{2k}]$, and $[\beta_k,\gamma_k] = [\alpha_{2k},\alpha_{2k+1}]$. These paired variables can be thought of as elements of $Z_N \times Z_N$. The tensors are then paired as   
\begin{align}
&B_m([\beta_k,\gamma_k])_{[\sigma_k,\tau_k],[\sigma_{k+1},\tau_{k+1}]} = \\ \notag 
&\delta_{\beta_k,\sigma_{k+1}-\sigma_k}  \delta_{-\gamma_k,\tau_{k+1}-\tau_k}  \nu_m(1,[\sigma_k,\tau_k],[\sigma_{k+1},\tau_{k+1}]) = \\ \notag
&\delta_{\beta_k,\sigma_{k+1}-\sigma_k}  \delta_{\gamma_k,\tau_{k+1}-\tau_k} \nu_{-m}(1,[\sigma_k,\tau_k],[\sigma_{k+1},\tau_{k+1}]), \rule[-1.5em]{0pt}{0pt}
\end{align}
where in the first line we used the definition of $\nu_m$ appearing in Eq. (\ref{Eq:Wen1DState}), and in the second line we used the fact that flipping the sign of just the $\tau$'s is equivalent to flipping the sign of $m$. As a result, our state appears as an MPS with site dimension ($d=N^2$) and bond dimension ($D=N^2$)
\begin{align}
\label{Eq:MPS1D}
\langle \{ [\beta,\gamma] \} |\psi_m\rangle &= Tr \left[ \Pi_k B_m([\beta_k,\gamma_k])\right].
\end{align}
Note that in App. (\ref{App:AKLT}) we show the relation between the above MPS for $D=Z_2 \times Z_2$ and the AKLT state. 

To use the classification of Ref. (\onlinecite{Schuch2011}), the symmetries of the MPS have to be identified. As one can verify (see also App. \ref{App:All}), $B_m([\beta,\gamma])$'s furnish a projective unitary representation of $Z_N \times Z_N$ 
\begin{align}
B_m(g) B_m(g') &= B_m(gg') \omega_{-m}(g,g') \\ \nonumber
\omega_m(g,g') &= \nu_m(1,g,gg'),
\end{align}
with $g,g' \in Z_N \times Z_N$. Let us associate with each element $g \in Z_N \times Z_N$, the matrix $B_m(g)$. Consider conjugating each matrix in the MPS by $B_m(g)$. Notably this operation does not change the trace in Eq. (\ref{Eq:MPS1D}) and thus leaves the state invariant. It can be presented as a unitary operation on Hilbert space through the following construction. Note that $B_m(g) B_m([\beta,\gamma]) B_m^{\dagger}(g) = \frac{\omega_{-m}(g,[\beta,\gamma])}{\omega_{-m}([\beta,\gamma],g)} B_m([\beta,\gamma])$. Interestingly $\frac{\omega_{-m}(g,[\beta,\gamma])}{\omega_{-m}([\beta,\gamma],g)}$ is equal to the character of $-mg$ in the representation labelled by $[\beta,\gamma] \in Z_N \times Z_N$ \cite{transgression}. In our notations, this character is $\chi_{-m g_l }(\beta) \chi_{-m g_r}(\gamma)$, where $g = [g_l,g_r]$ ($g_l, g_r \in Z_N$). Thus, conjugation by $B_m(g)$ is equivalent to acting on the state with the following unitary operator 
\begin{align}
O_g &= \sum_{ \{ [\beta,\gamma] \}}  |\{ [\beta,\gamma] \} \rangle \Pi_k \chi_{-m g_l }(\beta_k) \chi_{-m g_r}(\gamma_k) \langle \{ [\beta,\gamma] \}|.
\end{align}
By construction $O_g$ are symmetries of the MPS, and one may verify that they form a (non-projective) representation of $Z_N \times Z_N$. Alternatively stated, $|\psi_m\rangle$ possesses a $Z_N \times Z_N$ symmetry realized by the operators $O_g$. 

As shown in Ref. \cite{Schuch2011}, the topological phase of a MPS with a symmetry $Z_N \times Z_N$, can be determined by the projective representation of the matrices which implement the symmetry using conjugation. These matrices here are simply the $B_m([\beta,\gamma])$'s themselves. Since the set $\{ \nu_m \}_{m=1}^N$ spans ${\rm H}^2(Z_N \times Z_N,U(1))$, the different $B_m$'s span all projective representation and hence all SPTs in this symmetry class \cite{Schuch2012}.

In App. (\ref{App:SelfDuality}) we also show that an onsite unitary transformation ($\hat{F}$) which rotates between the symmetry-charge basis and the regular basis of Ref. \cite{Chen2011}, gives a mapping between the two sets of states 
\begin{align}
\label{Eq:Fourier}
\hat{F} | \tilde{\psi}_{-m}\rangle = {\rm \hat{d}_1}|\tilde{\psi}_m\rangle = |\psi_m \rangle. 
\end{align} 
The above relation also implies a non-local symmetry of our states and $|\tilde{\psi}_m\rangle$ of Ref. \cite{Chen2011}, given by ${\rm \hat{d}_1}$ times a local change of basis ($\hat{F}$). 

%

\section{Two dimensions}

We turn to discuss the two dimensional case. We again solve Eq. (\ref{Eq:Consistency}), only this time on a triangular lattice with both $\alpha_i$ and $\phi_i$ sitting on sites (vertices). We focus on a $Z_2$ symmetry, where the solution is quite simple and given by 
\begin{align}
\label{Eq:LatticeFlux}
{\rm d}_2[ \{ \phi \}]_j &= \frac{ \left \langle \sum_{\langle i,j\rangle} \langle \phi_i - \phi_{i-1} \rangle_{Z_2} \right \rangle_{Z_{2^2}} }{2} + a_0,
\end{align} 
where $\phi_i$ are arranged in clockwise order, $\langle ... \rangle_{Z_2}$ means taking modulo $2$, $a_0 \in \{ 0,1\}$ is an arbitrary number which we shall later fix. 

\begin{figure}[h!]\vspace{-0.2cm}
\includegraphics[width=93mm, trim = 140 350 90 150, clip=true]{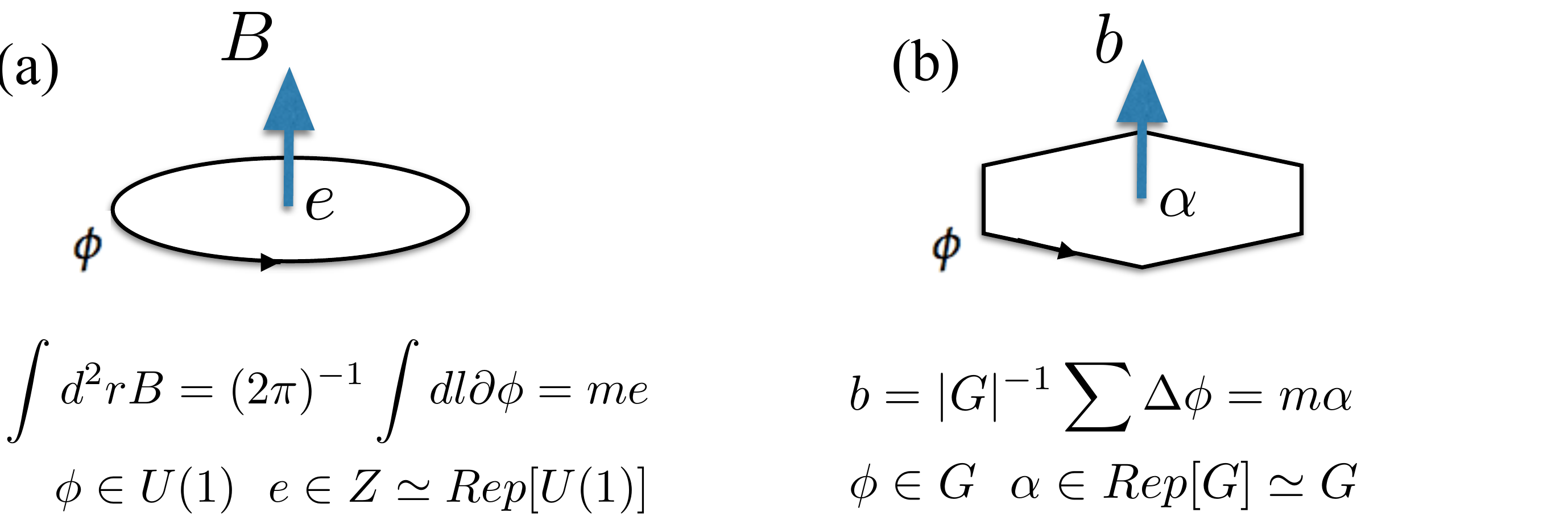}\vspace{-0.2cm}
\caption{\small (a) In the Laughlin state, each electron has a $U(1)$ charge ($e$) and also carries a magnetic flux $B=m e/\hbar$. (b) In the SPTs described here, each of the on-site degrees of freedom carries a $G$ charge, $\alpha$, and also a symmetry-flux, $b$. In both cases charges correspond to representations ($Rep$) of the symmetry group. These fluxes are the vorticity of $\phi$ which is either a $U(1)$ field (a) or a discrete field taking values in $G$ (b).}
\label{Fig:Flux}
\end{figure}

Despite appearances Eq. (\ref{Eq:LatticeFlux}) is a natural algebraic definition of vorticity (see Fig. (\ref{Fig:Flux})). To show this consider replacing $2$ by $N$ in the above, and taking the limit of large $N$. By presenting $Z_N$ as points on the complex circle, we asymptotically obtain the $U(1)$ group. A $U(1)$ singled valued function defined on a disk with a hole, may show any integer vorticity. A way of locally measuring this vorticity is the following:  (a) take a differential of the $U(1)$ phase $\partial \phi dl$ (b) lift it to be a variable in ${\mathbb R}$ within $[-\epsilon,\epsilon]$ (c) integrate the differential around the circle and (d) divide the result by the periodicity of $U(1)$ to obtain a number in ${\mathbb Z}$ equal to the vorticity. This procedure is well-defined because $U(1)={\mathbb R}/{\mathbb Z}$. In the above equations we do the same only interchange ${\mathbb R}$ by $Z_{N^2}$, integration by a discrete sum and $\epsilon$ by $1/N$. This procedure is well defined since $Z_N = Z_{N^2}/Z_N$. We note that our discrete vorticity is closely related to the Bockstein homomorphism  \cite{Hatcher2002,Ryan2014} in singular cohomology (or lattice gauge theory) where it is used for lifting a 1-cocyle (gauge field) to a 2-cocycle (curvature). 

\begin{figure}[h!]\vspace{-0.2cm}
\includegraphics[width=80mm, trim = 20 250 500 230, clip=true]{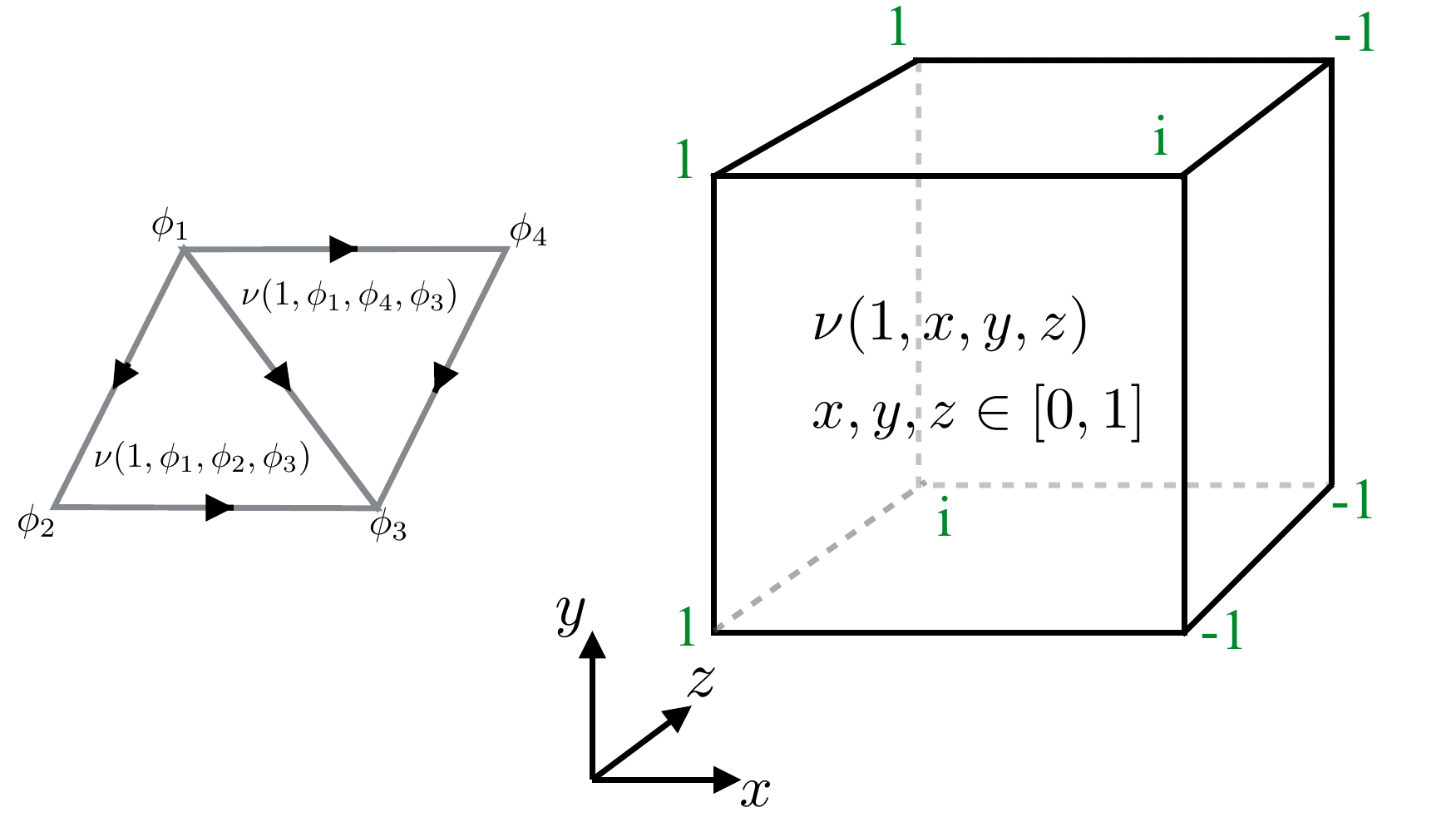}\vspace{-0.2cm}
\caption{\small The box depicts the values of $\nu(1,x,y,z)$ appearing in Eq. (\ref{Eq:Tensor2D}). Considering a triangular lattice with Ising spins ($\phi_i$) taking the values $0$ and $1$, we associate such a factor with each triangle and order the three arguments according to the branching structure. The product of all such local factors yields $(-1)$ to the number of domain walls in the spin configurations, a global quantity.}
\label{Fig:Box}
\end{figure}

We proceed by analyzing the $N=2$ case and leave a more generic study for future work. Given ${\rm d}_2$ with $a_0 = 1$, and Eq. (\ref{Eq:FluxAttachment}), the overall phase factor is determined and given by 
\begin{align}
\label{Eq:VorticityDomain}
e^{i \Theta_m(\{ \phi \})} &=(-1)^{\#dw[\{\phi\}]} 
\end{align}
where $\# dw[\{ \phi \}]$ denotes the number of domain walls in the $\{\phi\}$ configuration. To verify the above note that in the case where all surrounding $\phi$'s around a site $j_0$ are equal, flipping $\phi_{j_0}$ (i.e. adding $1$ modulo $2$) either creates or destroys a domain wall in the $\phi$ variables and so changes the domain wall number parity and the above right hand side receives a minus sign. Consistently with Eq. (\ref{Eq:FluxAttachment}), the discrete vorticity in this case is $a_0 = 1$ and $\chi_{m=1}(a_0) = -1$.  Similarly one can check all other $\phi$ combinations around the hexagon surrounding $j_0$. Interestingly, this overall phase factor can be divided into product of local factors thereby allowing a local TPS representation. The TPS is then given by 
\begin{align}
\label{Eq:Tensor2D}
|\psi_{Z_2}\rangle &= \sum_{{\alpha},\{ \phi \}} \Pi_j A_j(\alpha_j) | \{ \alpha \} \rangle \\ \nonumber
A_{\phi_{\langle \cdot ,j\rangle}}(\alpha_j) &= \delta_{{\rm d}_2 [\{ \phi \}]_j,\alpha_j} \\ \nonumber &\times \nu(1,\phi_{j},\phi_{j+1},\phi_{j+3}) \nu(1,\phi_{j},\phi_{j+2},\phi_{j+3})
\end{align} 
where $\phi_{j},\phi_{j+1},..,\phi_{j+3}$ denotes the four vertices on the two triangles to the right of $j$ in anti-clockwise order. The values of $\nu(1,\phi_1,\phi_2,\phi_3)$ are detailed in Fig. (\ref{Fig:Box}). One may verify that they indeed count the domain wall parity. Unlike in 1D, the $\nu$ we obtained here is not a cocycle. 


Similarly to the 1D case, an algebraic correspondence based ${\rm d_2}$, can be established between our state and a known Ising SPT groundstate \cite{Levin2012}. The ground state of the latter is written in the symmetry-phase basis and given by
\begin{align}
\langle \{ \phi \} | \tilde{\psi}_{Z_2}\rangle &= (-1)^{\#dw[\{\phi\}]}.
\end{align}
It then follows that our TPS ($|\psi_{Z_2} \rangle$) obeys
\begin{align}
|\psi_{Z_2} \rangle &= {\rm \hat{d}_2} | \tilde{\psi}_{Z_2} \rangle, \\ \nonumber 
{\rm \hat{d}_2} &= \sum_{\{ \phi \}} | \{ {\rm d_2} [\{\phi\}] \} \rangle \langle \{ \phi \} |, 
\end{align}
where in order to relate the two vector spaces we associate the local $|\sigma_z\rangle$ basis on the right hand side with $|\alpha = (\sigma_z + 1)/2\rangle$ on the left hand side. Unfortunately the above relation is again short of implying that both wavefunctions are in the same phase. There is a further qualitative difference compared to $1D$, as ${\rm \hat{d}}_2$ is not invertible even with fixed boundary conditions. Notably, one can quantify this lack of invertibility of ${\rm \hat{d}}_2$ by measuring the entropy per spin of the charge-ensemble, obtained by acting with ${\rm \hat{d}}_2$ on an ensemble of random spins. Using brute-force numerics, we have calculated this entropy for 4x4,5x5 and 6x6 lattices, for randomized boundary conditions (the boundary condition are counted here as part of the lattice). Following this we find that the entropy per spin is $1$, $0.997(3)$, and $0.995(5)$ times that of a random spin. The bulk density of degrees of freedom projected out by ${\rm d}_2$ is thus very small.

To establish the topological nature of $|\psi_{Z_2}\rangle$ we begin with a few analytical observations. First note that dealing with a $Z_2$ symmetry in 2D, it is believed that there is only one possible nontrivial SPT, the Ising SPT \cite{Chen2011}. Thus any symmetry respecting, short range entangled state which is distinct from the simple paramagnet (a state with $\sigma_x = 1$ everywhere) must be in the Ising SPT universality class. Consider tri-partiting the lattice into $A$, $B$ and $C$ sublattices and placing it on periodic boundary conditions which are consistent with this lattice labeling. Conveniently, flipping (adding $1$ modulo $2$) all the $\phi$ variables on one sublattice, has the effect of flipping the $\alpha$ variables on the remaining sublattices only. As a result one can show that     
\begin{align}
|\psi_{Z_2} \rangle &= \Pi_{a \in A} \sigma^a_x \Pi_{j \in [B,C]} \sigma^j_z |\psi_{Z_2} \rangle,
\end{align}
where $\sigma^j_z$ flips $\alpha_j$, and $\sigma^j_x$ gives a factor $2 \alpha_j-1$. From this it follows that the average of $\sigma_x$ is zero, implying the our state is very different from the simplest trivial paramagnet in which this quantity is simply one. It can similarly be shown that average of $\sigma_z$ and also that of the phase operator of Ref. (\onlinecite{Levin2012}) given by $\Theta_j =\Pi_{\langle q,q',j\rangle}(i)^{\frac{1-\sigma^q_z \sigma^{q'}_z}{2}}$ (where $q',q,j$ denote the vertices of the triangles containing $j$) both vanish. All these expectation values match with $|\tilde{\psi}_{Z_2}\rangle$. Nonetheless we find that these are not the same wavefunctions: Calculating $\Theta_j \sigma^j_x$ numerically for 3x3,4x4 and 5x5 lattices with periodic boundary conditions, we find that it is $0.82(6),0.12(4)$ and $0.08(1)$ respectively, while it is strictly one for $|\tilde{\psi}_{Z_2}\rangle$. Last we note that in the bulk, $\phi$ flips can only create/annihilate $\alpha=1$ sites in pairs. Thus in direct analogy with regular vorticity, in periodic boundary conditions there cannot be an overall odd vorticity (charge). Since the global Ising symmetry acts here as $(-1)$ to the number of nonzero charges, $|\psi_{Z_2}\rangle$ obeys the global Ising symmetry. In particular this implies that any product of an odd number of $\sigma_z$ or $\sigma_y$ averages to zero. 

\begin{figure}[h!]\vspace{-0.2cm}
\includegraphics[width=123mm, trim = 70 350 260 0, clip=true]{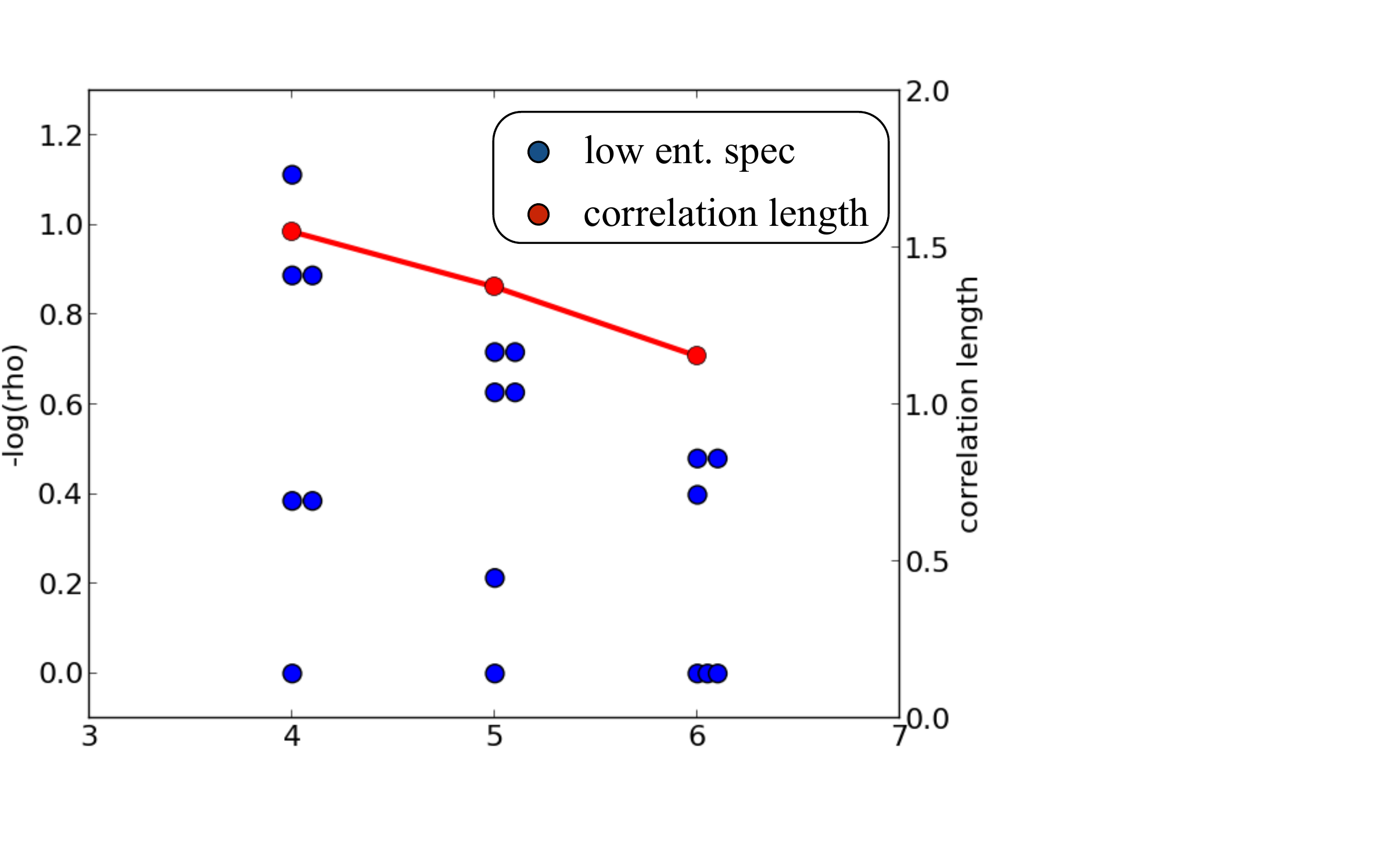}\vspace{-0.2cm}
\caption{\small Correlation length and sixth lowest entanglement spectrum eigenvalues for the TPS in Eq. (\ref{Eq:Tensor2D}) on an infinite cylinder of circumference 4,5 and 6 lattice sites. The TPS shows a small and decreasing correlation and are consistent with a gapless entanglement spectrum. }
\label{Fig:Results}
\end{figure}

To proceed, we use a variation of the PEPS algorithm \cite{Verstraete2004} suitable for our TPS  (see App. (\ref{App:PEPS}) for details). We obtain the spectrum and eigenvalues of the transfer operator associated with the TPS and its transpose, on a cylindrical geometry of circumferences 4,5, and 6 sites. Using this data we bound the correlation length and obtain the entanglement spectrum for a half infinite cylinder \cite{Schuch2011}. The results for the TPS are shown in Fig. (\ref{Fig:Results}). The correlation length is close to one and decreases with increasing system size. This justifies our finite size method and strongly suggests that the TPS describes a non-critical state (note that unlike MPSs, finite bond dimension TPSs may also describe critical states \cite{Perez-Garcia2007}). The entanglement spectrum, showing the six smallest eigenvalues, appears to be non-gapped.

To study properties of the TPS on larger systems we use a Monte-Carlo approach and evaluate observables by sampling the $\alpha$ (charge) configurations appearing in the TPS randomly. The seemingly difficult computation here is to find the amplitude of each sampled charge configuration, or equivalently, to invert ${\rm d}_2$. However due to the almost one to one nature of the ${\rm d}_2$ mapping, this task can be carried out rather efficiently. To this end we use a tree-search algorithm which calculates the source $\phi$  configurations by gradually iterating over 1D slices of the charge configuration. An efficient geometry to do this source search, is a cylindrical one with an initial $\phi=0$ boundary conditions on the two initial lines. These boundary conditions mimic a paramagnet, since constant $\phi$ yields a constant charge configuration. Given the charge configuration on the starting 1D slice, one calculates all possible lines of $\phi$'s which together with the two previous lines of $\phi$'s, agree with this charge configuration in terms of ${\rm d_2}$. For each such line of $\phi$'s the process repeats iteratively until the last slice is reached. Due to the low bulk entropy of sourceless charge configurations, we find that the recursion here branches rather slowly. On the last slice, we do not enforce any conditions on $\phi$, and therefore allow any 1D slice of charges. Since charges can fluctuate freely on this last slice this mimics a symmetry breaking or ferromagnetic boundary condition.  

\begin{figure}[h!]\vspace{-0.2cm}
\includegraphics[width=87mm, trim = 0 50 0 20, clip=true]{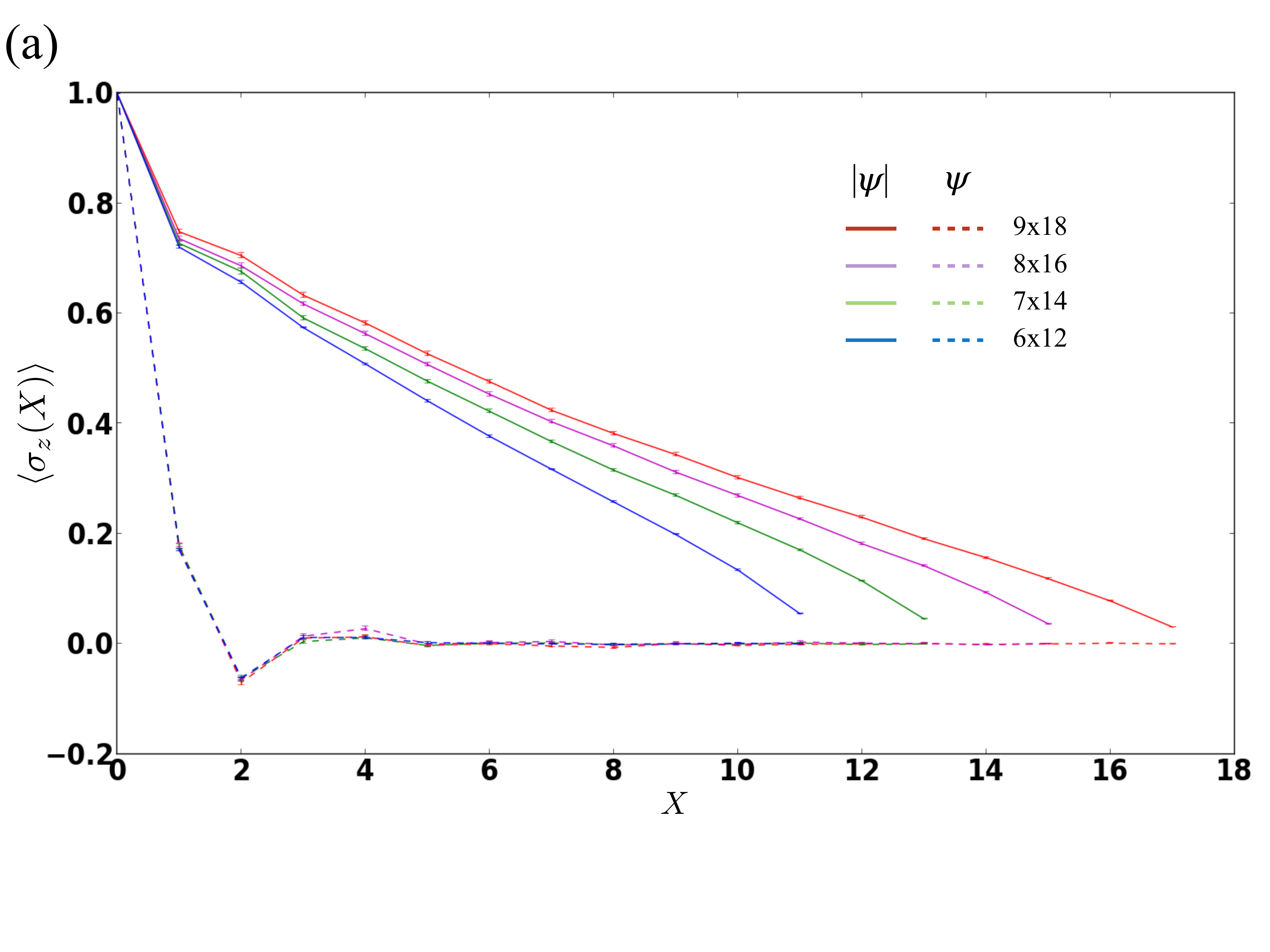}\vspace{-0.2cm}
\includegraphics[width=87mm, trim = 0 50 0 20, clip=true]{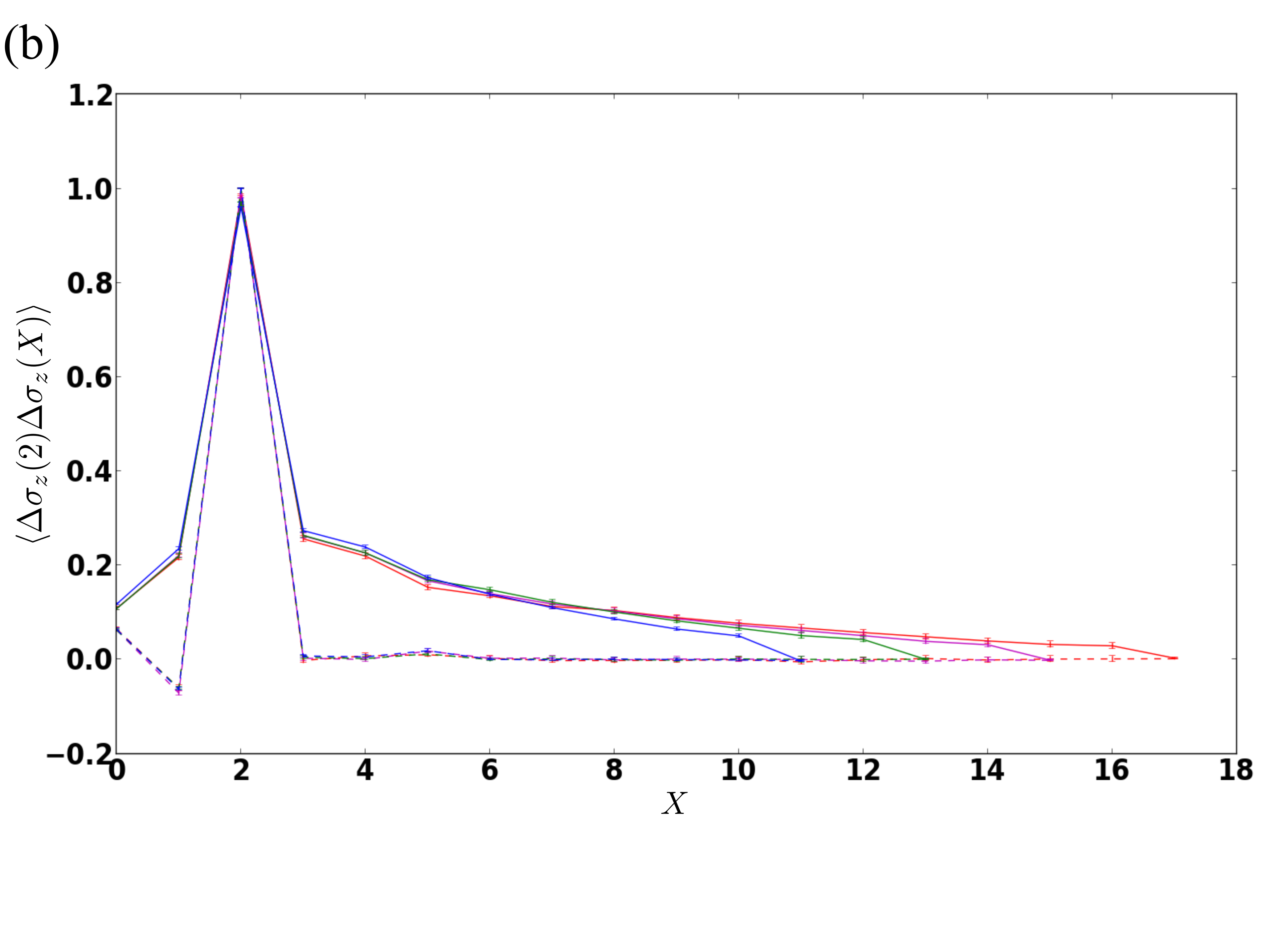}\vspace{-0.2cm}
\includegraphics[width=87mm, trim = 0 50 0 10, clip=true]{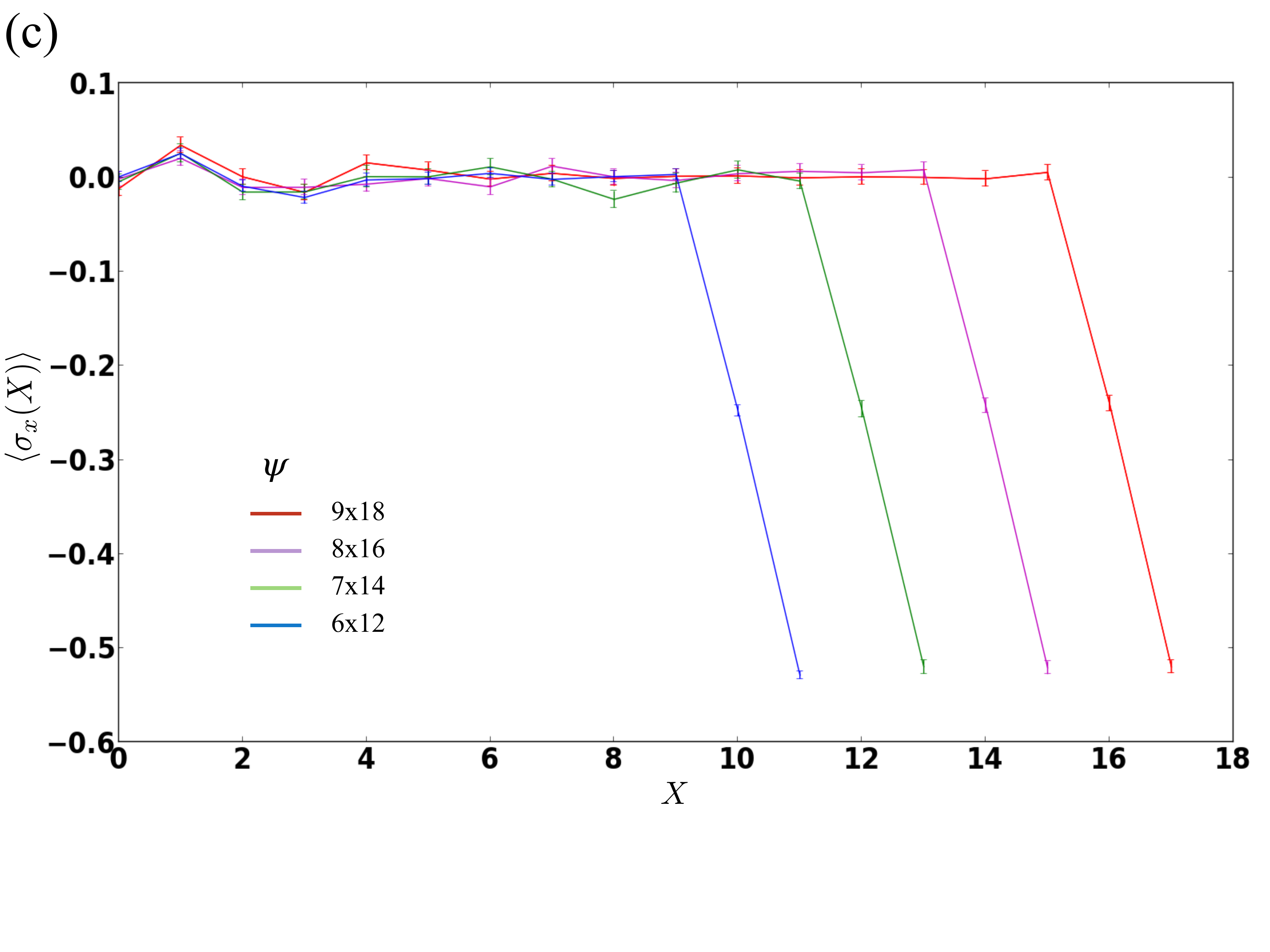}\vspace{-0.2cm}
\caption{\small Averages and correlations of $\sigma_z$ (panel (a) and (b)) as well as averages of $\sigma_x$ (panel (c)) along cylinders of different size with ferromagnet and paramagnetic boundary conditions on the left and right respectively. While the Laughlin-like wavefunction for the Ising SPT shows a fast decay of both boundary effects (panel (a)) and correlation effects (panel (b)), the modulus of the wavefunction reveals a hidden power law decay in both cases. The average of $\sigma_x$ is unaffected by the modulus transformation and shows a fast decay. }
\label{Fig:MCSZ}
\end{figure}

Using the above approach we have calculated various expectation values in cylindrical geometries of size $6\times12,7\times14,8\times16$ and $9\times18$. The triangular lattice was taken as a square lattice with an extra diagonal bond. The boundary conditions were, physically, a ferromagnet on the left end and a paramagnet in the other, in the sense described above. Fig. (\ref{Fig:MCSZ}), panel (a) depicts the average of $\langle \sigma_z(X) \rangle$ along the cylinder for both $|\psi_{Z_2}\rangle$ (solid line) and the modulus of $|\psi_{Z_2}\rangle$ (dashed line). While the effect of the ferromagnetic boundary quickly decays for $|\psi_{Z_2}\rangle$, the modulus shows strong evidence for a power law decay. Similar behavior appears in panel (b), when evaluating $\langle \Delta \sigma_z(2) \Delta \sigma_z(X) \rangle$. Panel (c) shows a fast decay of the average charge $\langle \sigma_x(X)\rangle$, for either $|\psi_{Z_2}\rangle$ or its modulus, as the two are equivalent in this case. We have also calculated averages and correlations of $\sigma_y(X)$ which appear to be zero everywhere. All errors bars show a $1\sigma$ ($~68\%$) confidence interval. The amount of charge configurations sampled is of the order of $1E+8$.  The above results strongly suggest that quasi-long-range order associated with a spontaneous breaking of the Ising symmetry exists in the modulo of the wavefunction. A similar behavior occurs in Laughlin's wavefunction \cite{Girvin1987} further strengthening the analogy between the two. As a comment on computational resources, we note that the $9\times 18$ results appearing in Fig. (a) took 6631 CPU hours and less than 2Mb per CPU. Execution time scaled roughly as 6 to the width. 

Lastly, we note that according to Ref. \cite{Cenke2014}, a $Z_2$ SPT should exhibits long range correlation of the type $C_{P}(x) = \langle P | \sigma_z(0) \sigma_z(x) | SPT \rangle / \langle P | SPT \rangle$ (so called strange correlators), where $|P\rangle$ is a topologically trivial state. Taking $|P\rangle$ to be a product of charge states $|\{ \alpha_i \}\rangle$, one finds that 
\begin{align}
&\langle |\psi_{Z_2}| | \sigma_z(0) \sigma_z(x) | |\psi_{Z_2}|\rangle \\ \nonumber &= \sum_{\{\alpha\}} \langle |\psi_{Z_2}| | \{ \alpha \} \rangle \langle \{ \alpha \} |\sigma_z(0) \sigma_z(x) || \psi_{Z_2}| \rangle \\ \nonumber 
&= \sum_{\{\alpha\}} \left|\langle |\psi_{Z_2}| | \{ \alpha \}\rangle \right|^2 |C_{P=\{\alpha\}}(x)|,
\end{align}
where in the second equality we used the fact that both $\langle |\psi_{Z_2}| | \{ \alpha \} \rangle$, and $\langle \{ \alpha \} |\sigma_z(0) \sigma_z(x) | |\psi_{Z_2}| \rangle$ are real and non-negative. As result we find that when the correlation length of the strange correlators is bounded, the modulo of the wavefunction cannot contain quasi-long-range order. Consequently quasi-long-range order in the modulo of $|\psi_{Z_2}\rangle$, implies that $|\psi_{Z_2}\rangle$ itself has long ranged strange correlators and so it passes this test for topology as well. Interestingly, similarly to the quasi-long-range order implied by the numerics, the strange correlators obtained for the Levin-Gu wavefunction also show quasi-long-range-order and are related to a non-unitary CFT with central charge $c=-7$ \cite{Cenke2014}.  

\section{Summary}

Motivated by the Laughlin wavefunction, we have written down an ansatz for a TPS describing abelian, bosonic SPTs in 1D and 2D. We tested our ansatz and established its validity for 1D SPTs with $G=Z_N \times Z_N$. This was carried out either by rewriting it as an MPS and using the known classification \cite{Schuch2011} or by directly mapping it, through a local basis change, to the wavefunctions obtained using the group cohomology approach \cite{Chen2011}. Furthermore, a unitary transformation which maps these SPT phases onto broken symmetry states (a disentangler \cite{Else2013,Quella2013}) was read-off the phase factor of these states. 

In 2D we used the same ansatz to derive a TPS which we conjecture to be in the Ising SPT phase. As evidence, we found that in periodic boundary conditions it is an Ising symmetry respecting state with strictly short range correlations. It has strongly fluctuating symmetry charges ($\sigma_x$) with zero average and thus it is markedly different from the representative wavefunction of the trivial phase (a simple product state of $\sigma_x=1$ on every site). Taking the modulus of the TPS revealed a critical state which is a feature of several other topological phases \cite{Girvin1987,Read1989,Else2013,Quella2013}. Furthermore based on the conjecture that the entanglement spectrum and edge spectrum shared the same topological features \cite{Li2008,Fidkowski2010}, we may relate its non-gapped entanglement spectrum to a non-gapped edge spectrum, as one expects from an SPT phase. Lastly, the criticalness of the modulus implies the existence of a long range strange correlator \cite{Cenke2014}, which is also a characteristic of an SPT phase. 

Discrete duality transforms, such as site to bond duality in the $1D$ and the discrete vorticity transform in $2D$, ended up playing important roles in our construction. In particular all the states we obtained could be written as such transforms acting on known SPT ground states. It thus appears that these transformations take SPT states written in the symmetry-phase basis and yield SPT phases presented on the symmetry-charge basis.
In this sense the SPT phases we studied appear to be self-dual: The phase is preserved under a duality transformation times a local change of basis.  For the $1D$ case this self-duality symmetry was established rigorously. Considering acting with these duality transformation on other phases, one finds that paramagnetic states (in the symmetry-phase basis) map onto ferromagnetic states (in the symmetry-charge basis) while ferromagnetic states (in the symmetry-phase basis) map onto paramagnetic states (in the symmetry-charge basis).   

Our approach, which is a hybrid of Laughlin's approach and TPSs, has several advantages. It is a constructive microscopic approach, derived directly from the physical concept of composite particles and flux attachment. This concept has proven extremely useful in past studies of bosonic, fermionic, integer, and fractional quantum Hall states. One may therefore hope that its adaptation to SPTs will provide for similar generalizations and help explore these phases away from the well mapped realm of integer (short ranged entangled) bosonic SPTs. Furthermore having a TPS representation for the groundstate allows the use of efficient numerical techniques \cite{Schuch2012} and also derivation of parent Hamiltonians \cite{Perez-Garcia2007}. Lastly, this approach touches upon the intriguing correspondence between topological phases and broken symmetry phases \cite{Girvin1987,Read1989,Else2013,Quella2013} as well as the correspondence between $1+1$ CFTs and $2+1$ topological phases \cite{Moore1991}. 

{\it Acknowledgments:} Z. R. would like to thank Curt Von Keyserlingk, Ryan Thorngren, Michael Zaletel, Jer\^{o}me Dubail, John Chalker, Bruce Bartlett, Thomas Scaffidi, and the referees of PRX for helpful comments and discussions. This work was supported by EPSRC. 

\appendix 

\section{Detailed derivation of a finite-size Laughlin TPS}
\label{App:LaughlinTPS}
 
Here we derive a finite-size TPS representation of the Laughlin state. Consider a bosonic Laughlin wavefunction at $1/m$ filling with $n$ particles in a droplet of radius of approximately $R=\sqrt{2(n-1)m}$, in units of the magnetic length which we set to unity     
\begin{align}
\Psi_{m}(z_1,...,z_n) &= \Pi_{i<j} (z_i - z_j)^{m} e^{ -\sum_i \frac{1}{4}|z_i|^2}, 
\end{align}
also consider its discrete version ($\tilde{\Psi}_{m}$) where  $z_i$ take values only on a lattice of resolution $\epsilon \rightarrow 0$. Note that there is no obstruction to discretizing space under a Laughlin state. In fact there are lattice models for which the discrete Laughlin wave function is an exact, topologically distinct, ground state \cite{Kapit2010,Scaffidi2014}. Here we show that one can find a tensor product state ($|TPS_m \rangle$) with continuous indices such that the overlap $\langle \tilde{\Psi}_m | TPS_m \rangle > 1-O(\epsilon/r_0)$,  provided that one keeps $\epsilon \ll R/n \propto R^{-1}$. 

To this end we introduce a $U(1)$ scalar field $\phi \in [0..1]$ which lives on the lattice and use it to decouple the flux from the particles as follows  
\begin{align}
\label{AppEq:TPS1Q}
\tilde{\psi}_{m}(z_1,...,z_n) &= C_0 \int{\rm D} \phi e^{ -\int d^2 r \frac{\beta}{2} (\nabla\phi)^2 + i  \pi m (\phi \rho(z)-\rho_0)} \Pi_i A_i \\ \nonumber 
A_i &= \delta \left(\int_{O_{z_i}} dl \nabla \phi - 1 \right), 
\end{align}
where $C_0$ is a normalization constant, $\beta=\pi m$, $\rho(z) = \sum_{i} \delta(z-z_i)$, $\rho_0$ is the neutralizing background charge, and $O_{z_i}$ is an $r_0$-circle around $z_i$, and derivatives and spatial delta functions should be thought of as their lattice counterparts. The boundary conditions for $\phi(r >> R)$ are $(2\pi)^{-1}n\Im \log(z)$.  We note that $\beta > \beta_c \approx 8 \pi \cdot 1.12$, the scalar field theory is on the ordered side of the Kosterlitz-Thouless transition \cite{Kosterlitz1973}. 
%

Let us be concrete about the lattice version of the vorticity operator ($\int_{O_{z_i}} dl \nabla \phi$). As a first attempt at defining it, one may consider the $r_0$ radius of integration, perform the discrete derivative, lift it from $U(1)$ to $R$, and sum in $R$. Whenever the field is smooth the resulting derivative is $\epsilon-$close to unity and so there is no ambiguity in lifting it to $R$. When the field is not smooth, the discrete derivative may be of the order of $1$ and ambiguities arise. The impact of such singularities should be strongly suppressed by decreasing ($r_0/\epsilon$). The reason is that, as shown later, $\beta > \beta_c$ for $m>8$. The field-theory on the scale of $r_0$, can then be made arbitrarily close to the smooth Gaussian fixed-point \cite{Polyakov1987} by increasing $r_0/\epsilon$. If we measure the vorticity on a scale $r_0$, the most problematic (common) singularity will be vortex anti-vortex pairs appearing on the scale $r_0$. If the vortex is very close to some $z_i$, the vorticity in most of the $r_0$ region will appears to be $1$. Consequently if we define the vorticity operator on the scale $r_0$, by some simple averaging procedure, we would be misled to think that there is vorticity there, while actually there is none. The probability of such an event is exponentially suppressed in the energy of such a configuration which goes as $\frac{\beta}{4\pi} \log(r_0/\epsilon)$, and so it goes as $(\epsilon/r_0)^{\beta/4\pi}$. Later on we will see that we actually need to take $r_0/\epsilon$ to scale as $R^{-1}$, and so this is not an important restriction.  

We turn to establish the validity of Eq. (\ref{AppEq:TPS1Q}). Note that the background charge term can be removed by transforming $\phi \rightarrow \phi + i \frac{\rho_0}{2} z^2$, while yielding the desired Gaussian factors of the Laughlin state. Next the delta function constraint can be removed by transforming $\phi \rightarrow \phi + \phi_0$ where $\phi_0 = \sum_i\frac{1}{2\pi} \Im \log(z-z_i)$.  Consequently the exponent of the $\phi$ field ($-S[\phi]$) changes to $S[\phi]+S_0[\phi_0,\phi]$ where
\begin{align}
\label{EqApp:S0}
S_0[\phi_0,\phi] &= \int d^2 r \frac{\beta}{2} (\nabla \phi_0)^2 + \beta \nabla \phi \nabla \phi_0 + i \pi m \phi_0 \rho(z). 
\end{align}
Note that boundary conditions of $\phi$ are now constant, and therefore allow no net vorticity.  
The last term on the above right hand side is given by 
\begin{align}
&\int d^2 r i \pi m \phi_0 \rho(z) = \sum_{i,j} i \frac{m}{2} \Im \log(z_i - z_j) \\ \nonumber 
&= \sum_{i<j} i m \left [ \Im \log(z_i-z_j) + \frac{1}{2} \Im \log(-1)\right] + \sum_{i} \frac{i m}{2}\Im \log(0).  
\end{align}
Out of the resulting three terms, the second is just $\pi i mN(N-1)/4$ and can be absorbed into $C_0$. To define the third term we require a lattice definition of the $\phi_0$ shift. We choose it such that $\log(0)=0$. 

Next we discuss the second term on the right hand side of Eq. (\ref{EqApp:S0}). We use integration by parts to obtain $\phi \nabla^2 \phi_0$ along with a vanishing boundary term. Away from the singularities of $\phi_0$ we may use the continuum limit where $\nabla^2 \phi_0 = 0$. In a few $\epsilon$ radius around the singularities of $\phi_0$, a more delicate analysis is required. Let us taylor expand $\phi$ around some particle ($z_j$) as $\phi(z+z_j) = \phi(z_j) + z \partial_z \phi(z_j)+ ..$. As one can check the $\phi(z_j) \nabla^2 \phi_0$ term vanishes exactly even on the lattice. Thinking of the underlaying lattice as rectangular, $\pi$ rotation symmetry also removes the next order term. Following this, contributions from the vortex cores will be smaller than $\sum_j \epsilon^2 \partial^2 \phi(z_j)$ and therefore negligible. 

We turn our attention to $\frac{\beta}{2} (\nabla \phi_0)^2$. Away from the singularities we may again use the continuum approximation with minor $\epsilon$ corrections of the type we had previously. By a $\pi/2$ rotation of $\nabla \phi_0$, we find that it describes the electric field ($E$) of a set of $2D$ coulomb charges at positions $z_i$ (or equivalently the electric field of a set of 1D wires parallel to a fictitious $z$ axis). The action is therefore $\int d^2 r E^2$ and proportional to the overall electrostatic energy of this field configuration. Rewriting $E = \nabla \varphi $, where $\varphi = \sum_i (2\pi)^{-1} \Re \log(z-z_i)$, we use integration by parts to get Coloumb's law for the charges ($\sum_{i,j} \frac{\beta}{4 \pi} \Re \log(z_i-z_j)$) along with a surface term which gives the infrared divergence proportional to $n^2\log(R)$. These infrared divergence can again be absorbed into the normalization, provided $\epsilon$ is small enough. Focusing on the Coulomb part, terms in this sum with equal $i$ and $j$, give rise to a some core energy term $E_{core} \propto 1/\epsilon^{-1}$. Since the vortex number is fixed here, their contribution is constant can be absorbed into $C_0$. Note that when $z_i$ and $z_j$ come $\epsilon$-close, the continuum description breaks down. Also before this, there may be running coupling effect in $\beta$. The length scale on which these effects start interfering is again controlled by $\epsilon$ and therefore we allow ourselves to ignore this.   

The last term which couples the $U(1)$ field to the particles, is the preexisting $\int i \pi m \rho(z) = i m \pi \sum_{i} \phi(z_i)$ term. Completing the square, this results in an additional vortex repulsion term ($ \sum_{i,j}\frac{\pi m^2}{4 \beta} \Re \log(z_i-z_j)$). The $\phi$ field then becomes decoupled and can be integrated out leading to yet another change in $C_0$. 

Finally we choose $\beta$, such that $\frac{\pi m^2}{2 \beta} + \frac{\beta}{2 \pi} = m$ so that the coefficients of the real and imaginary part of the log are both equal to $m$. This yields a single solution which is ($\beta = \pi m$) such that the desired $\sum_{i<j} m\log(z_i-z_j)$ term appears in the exponent. Notably, $\beta > \beta_c$ for $m>8$.

Last we recast Eq. (\ref{AppEq:TPS1Q}) into tensor product form. Instead of describing the position of each particle we work with a discrete density field, $\rho_{ab}$, and describe the particle number at each position. The resulting expression is   
\begin{align}
\label{AppEq:LaughlinTPS}
\langle \{ \rho_{ab} \} | TPS_{m} \rangle &= \int \Pi_{ab} [d \phi_{ab}]  A_{\phi}(\rho_{ab}) \\ \nonumber 
A_{\phi}(\rho_{ab}) &= \delta \left(\int_{O_{z_{ab}}} dl \nabla \phi  - \rho_{ab} \right) e^{i \pi m \phi_{ab} (\rho_{ab}-\epsilon^2 \rho_0)} \\ \nonumber &\times e^{ -\frac{\pi m}{4} (\nabla \phi_{ab})^2} 
\end{align}  
where $|\{\rho_{ab}\}\rangle$ is a state in the occupation basis. The above expression now appears as a tensor product state with $\phi$ serving as the tensor indices and taking values in the group $U(1)$. 
Notably, as well as forcing vorticity at particle positions, the above form also forces zero vorticity everywhere else. As long as $\beta > \beta_c$ this should not matter much as net vorticity, on the scale $r_0$, is already exponentially suppressed. When $\beta < \beta_c$, a naive guess will be that these constraints suppress the Kosterlitz-Thouless transition.  

Last we discuss the main mode of failure of this Laughlin TPS in the thermodynamic limit. In the above, the scalar field is forced into a highly non-typical configuration with a finite density of positive vorticity. The circulation induced by this vorticity at radius $R$, will scale as $n$, the total number of particles. To make the continuum approximation valid at this radius we must require $R/\epsilon \gg n$ and so $\epsilon^{-1} \gg n/R \propto R$. Notably this mode of failure, becomes irrelevant once we consider a system with a discrete symmetry instead of a $U(1)$ symmetry. Then, there is no sense in thinking about positive vorticity, as charge/vorticity becomes a cyclic variable. Accordingly there is no notion of a macroscopic vorticity, which is the source of the problem here.   

\section{$\nu_m$ covers ${\rm H^2}(Z_N \times Z_N,U(1))$} 
\label{App:All}
Here we show that the cocycles used in Eq. (\ref{Eq:MPS1D}), fully cover the second cohomology group ${\rm H^2}(Z_N \times Z_N,U(1))$ in the sense that they give a representative cocycle in each equivalence class. 

From any such cocycle one can construct a projective representation of $Z_N \times Z_N$ as follows. Consider a vector space $|[\sigma,\tau]\rangle$, labeled by group elements and have the matrices $D([\sigma_0,\tau_0])$ act as $D([\sigma_0,\tau_0])|[\sigma,\tau]\rangle= |[\sigma+\sigma_0,\tau+\tau_0]\rangle$. This is gives the regular (and non-projective) representation of $Z_N \times Z_N$. Next we use $D$ and the cocycle to define a projective representation of $Z_N \times Z_N$ by
\begin{align}
[R([\sigma_0,\tau_0])]_{[\sigma,\tau],[\sigma',\tau']} &= [D([\sigma_0,\tau_0])]_{[\sigma,\tau],[\sigma',\tau']} \\ \nonumber &\times \nu_m(1,[\sigma,\tau],[\sigma',\tau'])
\end{align}
using the cocycle condition ($\nu_m(a,b,c)\cdot \nu_m(a,c,d) \cdot \nu^{-1}_m(b,c,d) \cdot \nu^{-1}_m(a,b,d) = 1$), one can verify that this is a projective representation and that 
\begin{align}
\label{Eq:RRep}
R([\sigma_0,\tau_0])R([\sigma_1,\tau_1]) &= R([\sigma_0+\sigma_1,\tau_0 + \tau_1])  \\ \nonumber &\times \omega_m([\sigma_0,\tau_0],[\sigma_1,\tau_1]), \\
\omega_m([\sigma,\tau],[\sigma',\tau']) \equiv &\nu_m(1,[\sigma,\tau],[\sigma'-\sigma,\tau'-\tau]) \\ \nonumber &= e^{-\frac{2 \pi i m}{N} \tau \sigma'}.
\end{align}
For any $m$ different than zero, the representation obtained is non-abelian and must therefore be projective. As a result, we find that all the cocycles are non-trivial. We further note that ${\rm H^2}(Z_N \times Z_N,U(1))=Z_N$ has a group structure under the action of multiplying cocycles. Since $\nu_m \nu^{-1}_{m'} = \nu_{ m-m'}$ results in a regular representation only when $m=m'$, we find that all cocycles are distinct. Having $N$ of them means they cover the entire cohomology group.

\section{Self duality in 1D SPTs}
\label{App:SelfDuality}
Here we construct a local, symmetry respecting, unitary transformation which maps between $|\tilde{\psi}_m\rangle$ and $|\psi_m\rangle$. To this end we limit ourselves to prime cyclic groups. This way multiplying $\phi$ by $m$ is an automorphism from the group to itself. Consider the local unitary Fourier transform, $[\hat{F}_m]_{\alpha,\phi} = \chi_{m \phi}(\alpha)$, which changes basis from the $\phi$ (symmetry-phase) basis and the $\alpha$ (symmetry-charge) basis. We argue that when acting on $| \tilde{\psi}_m\rangle$, this transform is equivalent to ${\rm d}_1$ up to complex conjugation. A useful tool in proving this is the following parent Hamiltonian for $|\tilde{\psi}_m\rangle$ 
\begin{align}
{\rm H_{sd}} = -\sum_i \sum_{p=0}^{N-1} e^{ \frac{ 2 \pi i m p}{N} (\tau_{i}-\tau_{i-1})} \sigma^{+p}_{i} +  e^{\frac{ 2 \pi i mp}{N} (\sigma_{i}-\sigma_{i+1})} \tau^{+p}_i,
\end{align}
where $[\sigma_i,\tau_i] = [\phi_{2i-1},\phi_{2i}]$, $\sigma^{+p}_i$ is a raise-$\sigma$-by-$p$-and-take-modulo-N operator (for example in the $N=2$ case, $\sigma^{+0} = I_{2x2}; \sigma^{+1} = \sigma_x$). Note that this Hamiltonian obeys two separate $Z_N$ symmetries, acting on the $\sigma$ and $\tau$ degrees of freedom with the regular action. 

Next we show that ${\rm H}_{sd}$ obeys a self-duality symmetry and that its groundstate is unique.  First recall the action of ${\rm \hat{d}}_1$ and $\hat{F}_m$ on the double site basis, spanned by all configurations of the form $|...,[\sigma_{i-1},\tau_{i-1}],[\sigma_i,\tau_i]...\rangle$ 
\begin{align}
&{\rm \hat{d}}_1 |...,[\sigma_{i-1},\tau_{i-1}],[\sigma_i,\tau_i],[\sigma_{i-1},\tau_{i-1}]...\rangle \\ \nonumber &=  |...,[...,\sigma_{i-1} - \sigma_i],[\tau_i-\tau_{i-1},\sigma_i - \sigma_{i+1}],[\tau_{i+1} - \tau_{i},...]...\rangle \\ 
&[\hat{F}_m]_{[\tilde{\sigma},\tilde{\tau}],[\sigma,\tau]} = \frac{1}{\sqrt{N}} e^{\frac{2 \pi i m(\tilde{\sigma} \sigma + \tilde{\tau} \tau)}{N}}.
\end{align} 
For interpreting ${\rm \hat{d}}_1$ as a symmetry, we need its inverse which, strictly speaking, does not exist. This is because any two symmetry-phase configurations which differ by a global rotation are mapped to the same state. Focusing on bulk effects, one can remedy this problem by fixing the left most two sites to be $[0,0]$.   

Conjugating $ {\rm H_{sd}}$ by ${\rm \hat{d}}_1$ we find  
\begin{align}
&{\rm \hat{d}} {\rm H_{sd}} {\rm \hat{d}}^{-1}_1 = -\sum_i [\sum_{p=0}^{N-1} e^{ \frac{ 2 \pi i m p}{N} (\sigma_{i})} \tau^{+p}_{i}\tau^{-p}_{i-1} \\ \nonumber & +  e^{ \frac{ 2 \pi i m p}{N} (\tau_{i})} \sigma^{+p}_{i} \sigma^{-p}_{i+1}],
\end{align}
Next using the fact that 
\begin{align}
\hat{F}_m^{\dagger} e^{ \frac{ 2 \pi i m p}{N} \tau} \hat{F}_m &= \tau^{+p} \\ 
\hat{F}_m^{\dagger} \tau^{+p} \hat{F}_m &= e^{- \frac{2\pi i m p \tau}{N}}, \\
\hat{F}_m^{\dagger} e^{ \frac{ 2 \pi i m p}{N} \sigma} \hat{F}_m &= \sigma^{+p} \\ 
\hat{F}_m^{\dagger} \sigma^{+p} \hat{F}_m &= e^{-\frac{2\pi i m p \sigma}{N}}, 
\end{align}
we find that 
\begin{align}
&\hat{F}_m^{\dagger} {\rm \hat{d}}_1 {\rm H_{sd}} {\rm \hat{d}}^{-1}_1 \hat{F}_m \\ \nonumber 
&= -\sum_i \sum_{p=0}^{N-1} e^{ \frac{ -2 \pi i m p}{N} (\tau_{i}-\tau_{i-1})} \sigma^{+p}_{i} +  e^{\frac{-2 \pi i mp}{N} (\sigma_{i}-\sigma_{i+1})} \tau^{+p}_i \\ \nonumber &= {\rm H}^{*}_{sd}
\end{align}

Next we wish to show that the ground state of this Hamiltonian is unique on closed boundary conditions. To this end we conjugate the Hamiltonian with the following unitary transformation 
\begin{align}
U &= | \{ [\sigma , \tau] \} \rangle \nu_m^{-1}(1,[\sigma_i,\tau_i],[\sigma_{i+1},\tau_{i+1}]) \langle  \{ [\sigma , \tau] \}|
\end{align}
As one can verify from Eq. (\ref{Eq:Wen1DState}), this transformation removes all phase factor from the wave function, leaving a state which is an equal superposition of all $\{ [\sigma,\tau] \}$ configurations. Similarly one finds 
\begin{align}
U^{\dagger} {\rm H_{sd}} U &= -\sum_i \sum_p \left[ \tau^{+p} + \sigma^{+p} \right].
\end{align}
Clearly the resulting Hamiltonian has a unique ground state and since $U$ is unitary, so does ${\rm H_{sd}}$. Thus when fixing $\phi$ on the boundary, the above Hamiltonian is gapped and has a unique ground state. It is then easy to show that   
\begin{align}
\label{Eq:Fourier}
\hat{F}_m | \tilde{\psi}_{-m}\rangle = {\rm \hat{d}_1}|\tilde{\psi}_m\rangle = |\psi_m \rangle, 
\end{align}
as advertised. Consistently with our initial definitions, the symmetry $G$ acts regularly on $|\tilde{\psi}_m\rangle$ and so, conjugated by $\hat{F}_m$, it should indeed act diagonally on our TPS. 

\section{TPS numerics}
\label{App:PEPS}
Here we study the correlation length and entanglement spectrum of the TPS in Eq. (\ref{Eq:Tensor2D}) both with and without the $(-1)^{\#dw[\{ \phi \}]}$ factor. We work in an infinite cylindrical geometry aligned along the $x$-direction. The underlying triangle lattice is skewed into a square lattice containing an extra diagonal bond on each square going in the right-up direction. To obtain the correlation length and entanglement spectrum, we use the transfer operator approach as described in Ref. \cite{Schuch2012}. Since our TPS differs slightly from standard PEPS, we re-derive some of their formulation. 

Consider two local operators placed at $x_1$ and $x_2$ given by 
\begin{align}
O_1 &= \sum_{\alpha_1,\alpha^{'}_1} c_{\alpha_1,\alpha^{'}_1} | \alpha_1 \rangle \langle \alpha^{'}_1 | \otimes I\\ \nonumber 
O_2 &= \sum_{\alpha_2,\alpha^{'}_2} c_{\alpha_2,\alpha^{'}_2} | \alpha_2 \rangle \langle \alpha^{'}_2 | \otimes I
\end{align}
where we make a slight abuse of notation and use the dummy indices to denote the position on the lattice. Calculating their joint expectation value involves tracing two conjugated TPSs 
\begin{align}
&\langle O_1 O_2 \rangle =  \\ \nonumber 
& \frac{\sum_{\{ \phi,\bar{\phi}, \alpha,\alpha^{'}\}} \Pi_i A_{\phi_{\langle \cdot, i\rangle}}(\alpha_i)  A^*_{\bar{\phi}_{\langle \cdot, j\rangle}}(\alpha^{'}_j)\langle \{ \alpha \}| O_1 O_2 | \{ \alpha^{'} \} \rangle }{\sum_{\{ \phi,\bar{\phi}, \alpha,\alpha^{'}\}} \Pi_i A_{\phi_{\langle \cdot, i\rangle}}(\alpha_i) A^*_{\bar{\phi}_{\langle \cdot, j\rangle}}(\alpha^{'}_j)\langle \{ \alpha \}| \{ \alpha^{'} \} \rangle},
\end{align}
The $A$ tensors are those used in the main text however due to technical reasons, the cocycle differs by an insignificant boundary term. More specifically we take $\nu'(1,a,b,c)$ which is $1$ ($i$) whenever the majority of $a,b$ and $c$ are zero (one) and add an additional $(-1)^{\phi_i}$ term. One can check that such factor also solves equation Eq. (\ref{Eq:Consistency}) with the previously used ${\rm d}_2$. For example if all $\phi$'s surrounding a site $j$ are zero, ${\rm d}_2[\{ \phi \}]_j = a_0 = 1$ and consequently a minus sign is expected. Accordingly, there will be no change in $\nu'$ in the six triangles around $j$ however due to the $(-1)^{\phi_j}$ factor there will be an overall minus sign. Since both this choice and the one used in the main text agree on periodic boundary conditions, they may only differ by a boundary term. Focusing on bulk properties one may ignore this subtlety. Indeed the boundary term can be absorbed into a unitary transformation of the transfer matrix defined below.

Let us first discuss how to use the transfer matrix method to efficiently calculate the above denominator. Consider dividing the cylinder into thin ring regions ($R$) whose width is two sites. The transfer operator of such a ring is defined as 
\begin{align}
[T]_{(\phi_l,\bar{\phi}_l), (\phi_r,\bar{\phi}_r)} &= \sum_{\{ \alpha \}_R} \Pi_{i \in R_r} A_{\phi_{i,L},\phi_{i,R}}(\alpha_i) A^*_{\bar{\phi}_{i,L},\bar{\phi}_{i,R}}(\alpha_i), 
\end{align}  
where $\phi_{i,L}$ and $\phi_{i,R}$ are the left and right subsets of the set $\{ \phi_{\langle \cdot,i \rangle},\phi_i \}$, since these two subsets overlap, a delta function forcing them to be equal is implicit in the above notation. The set $R_r$ denotes the indices on the right column of the region $R$. Using $T$, one can express $\langle \psi_{Z_2} | \psi_{Z_2} \rangle$ for a cylinder of length $N$ rings, as 
\begin{align}
\langle \psi_{Z_2} | \psi_{Z_2} \rangle &= \langle \chi_L| T^N | \chi_R \rangle, 
\end{align}
where $\langle \chi_L |$ and $ | \chi_R \rangle$ are determined by the boundary conditions. Applying similar ideas we can express the average as 
\begin{align}
\label{Eq:ExpectationViaTransfer}
\langle O_1 O_2 \rangle &= \frac{ \langle \chi_L | T^{n} M_1 T^{m} M_2 T^{N-m-n-2}| \chi_R \rangle} { \langle \chi_L | T^N | \chi_R \rangle}, 
\end{align}
with $M_k$ being 
\begin{align}
\label{Eq:MK}
&[M_k]_{(\phi_l,\bar{\phi}_l), (\phi_r,\bar{\phi}_r)} = \\ \nonumber &\sum_{\{ \alpha,\alpha^{'} \}_R} \Pi_{i \in R} A_{\phi_{i,L},\phi_{i,R}}(\alpha_i) A^*_{\bar{\phi}_{i,L},\bar{\phi}_{i,R}}(\alpha_i) [O_k]_{\{ \alpha \}_R, \{ \alpha^{'} \}_R}.
\end{align}

Next we discuss an important symmetry of $T$ which stems from the fact that flipping all the indices does not affect the charge configuration. As an initial guess for what the symmetry associated with this redundancy may be, consider flipping either the $\phi$ or $\bar{\phi}$ indices of $T$. As far as $\alpha$'s are concerned, this is clearly a symmetry. However the phase factor changes. Since $R$ contains two columns, the $\Pi_{i \in R} \phi_i$ is always symmetric under such a flip. However the $\nu$ factors change such that $1$ becomes $i$ and $i$ becomes $1$ (recall that $\nu(1,\phi_1,\phi_2,\phi_3)$ is $1$ ($i$) if the majority of $\phi$'s are $0$ ($1$)). Thus each triangle in $R$ gets multiplied by either $i$ or $-i$ depending on its majority spins. Since the number of triangles is even, this always results in a $\pm 1$ relative phase factor. For even circumference, this factor is simply $(-1)$ to the number of majority $1$ triangles. Conveniently, we find that this factor, which is naively a function of both the left and right indices of $T$, can be written as a product of two functions acting on the left and right indices of $T$ separately. As a result we obtain the following Ising-advanced ($I_A$) and Ising-retarded ($I_R$) symmetries of $T$ 
\begin{align}
[I_A]_{\{ \phi,\bar{\phi} \},\{ \phi^{'},\bar{\phi}^{'} \}} &= \delta_{ \{ \phi \}  = \{ -\phi^{'} \} } \delta_{ \{ \bar{\phi} \}  = \{ \bar{\phi}^{'} \} } (-1)^{ \# [11-bonds](\{ \phi^{'} \})}
\end{align}
where $\# [11-bonds](\phi^{'})$ denotes the number of adjacent sites whose $\phi$'s are both equal to $1$ on the right column of $\{ \phi^{'} \}$. The Ising-retarded symmetry is defined in exactly the same only with $\phi$ and $\bar{\phi}$ exchanged. 

Notably, the above two Ising symmetries are also symmetries of $M_1$ and $M_2$ and more generally of any such matrix representing an operator acting on the physical space. Indeed, physical operators such as the above $O_1$ and $O_2$, when presented in transfer matrix form, are mapped to different pairing of $A(\alpha_j)$ between the retarded and advance sector (see Eq. (\ref{Eq:MK})). To maintain the above lifted-Ising symmetries, one just requires the $A$ tensors within each sector to be function of $\phi$ fluxes and not $\phi$ themselves. Consequently these symmetries are automatically obeyed by any transfer matrix involved in calculating physical observables. It reflects a redundancy in our description rather than an actual physical symmetry.

To obtain the decay of correlations, we numerically obtain the six maximal eigenvalues of $T$ for cylinders of circumference $4,5$ and $6$. The results are 
\begin{align}
\lambda_4 = &[ 17.60(2),  16.67(7), 8.75(3),   \\ \nonumber & 8.40(0), 8.34(5),  -7.85(7)]\\
\lambda_5 = &[ 34.47(8),  34.31(4), 16.42(2),  \\ \nonumber & 4.40(5) +13.55(7)i, 14.30(0),   4.40(5)-13.55(7)i] \\ \nonumber
\lambda_6 = &[ 70.04(8), 69.33(6),-14.72(5)+25.50(5)i, \\ \nonumber  &-14.72(5)+25.50(5)i,-28.71(9),  29.45(1)]
\end{align} 
In both cases, the two maximal eigenvalues show a different $I_A$ and $I_R$ eigenvalues. The (marginally) larger of the two has a $+1$ eigenvalue for both $I_A$ and $I_R$ and the smaller one has a $-1$ eigenvalue for both symmetries. As we next show, based on symmetry consideration these two maximal eigenvalues cannot transmit any correlations. Taking this constraint into account, as well as Eq. (\ref{Eq:ExpectationViaTransfer}), the correlation length $\chi_n$ is given by $e^{-\chi_n} = \lambda_n[2]/\lambda_n[0]$ (as these are the two closest eigenvalues within the same sector), and we find $\chi_4 = 1.55(1), \chi_5 = 1.35(7), \chi_6 = 1.15(4)$. This short scale, suggests that the above results are already close to the thermodynamic limit.

Studying the TPS without the phase factor we find 
\begin{align}
\lambda_4 &= [ 19.54(2),16.88(5), 16.88(5), 11.72(8), \\ \nonumber & 10.15(2),  9.21(1)] \\
\lambda_5 &= [ 39.06(7),  33.89(1), 33.89(1),  24.99(4), \\ \nonumber & 21.73(0),  14.63(9) +3.24(2)i] \\
\lambda_6 &= [ 78.90(4),  69.70(5), 69.70(5), 53.56(4), \\ \nonumber & 48.92(2), -32.37(4)] 
\end{align}
Here the definition of the $I_A$ and $I_R$ symmetries are simply spin flips, as there is no phase factor. Checking the $I_A$ and $I_R$ eigenvalues we find that the largest eigenvector has both positive $(+,+)$, unlike before the two smaller ones are a mixture of $(+,+)$ and $(-,-)$. Consequently the symmetry does not protect mixing by physical operators. Indeed taking mixed boundary conditions containing the above three top eigenvectors, we find various values for the magnetization, typically of the order of $1/2$.  We also obtain large correlation lengths $\chi_4 = 6.84(2), \chi_5 = 7.03(5), \chi_6 = 8.06(7)$ suggesting a critical phase. 

We turn to show how symmetry protects the mixing of transfer matrix eigenvalues. Since $T$ is not Hermitian, in general it can only be brought to a Jordan form using a similarity transformation ($S^{-1} T S$). We define the right $|\lambda,i )$ and left $( \lambda,i |$ eigenvectors of $T$ obeying $T | \lambda,i ) = \lambda |\lambda ,i) + | \lambda,i-1)$ (where $|\lambda,0) \equiv 0$), and $( \lambda,i | T = \lambda  ( \lambda,i | + (\lambda,i+1|$ (where $(\lambda,N+1| \equiv 0$ where $N$ is the size of the block). These eigenvectors obey $(\lambda,i | \lambda',j ) = \delta_{\lambda,\lambda'} \delta_{ij}$. Given $[T,U]=0$, $T$ acts within each any eigenvalue  subspace of $U$, and so the Jordan blocks can be assigned with good $U$ eigenvalue numbers ($\chi_{\lambda}$). Consequently $U |\lambda,i) = \chi_{\lambda} | \lambda,i)$, and $(\lambda,i| U = (\lambda,i| \chi_{\lambda}$, and $(\lambda,i|U^{-1} = \bar{\chi}_{\lambda}$. Considering now a different matrix $M$, respecting the symmetry $U$. We find that $( \lambda | M | \lambda' ) = (\lambda | U^{-1} M U | \lambda') = \chi_{\lambda} \bar{\chi}_{\lambda'} ( \lambda | M | \lambda' )$. Consequently it vanishes when $|\lambda)$ and $|\lambda')$ have different $U$ eigenvalues. Inserting an $\sum_{\lambda} | \lambda )(\lambda|$ resolution of identity in Eq. (\ref{Eq:ExpectationViaTransfer}), before and after $M_1$ and $M_2$, one finds that the two maximal and almost degenerate eigenvalues of the TPS (with phase factors) are unable to transmit correlations. 

To obtain the entanglement spectrum we again follow Ref. \cite{Schuch2012}. For simplicity, we normalize $T$ such that its largest eigenvalue is exactly $1$. Consider the reduced density matrix of a very long cylinder cut in the middle. One can define the following two quantities capturing the ``quantum state'' of the unphysical indices at the cut 
\begin{align}
[\sigma_R]_{\{ \phi \}, \{ \bar{\phi} \}} &= \langle \{ \phi \}, \{ \bar{\phi} \} |T^{N/2} | \chi_R \rangle \\
[\sigma_L]_{\{ \phi \}, \{ \bar{\phi} \}} &= \langle \chi_L | T^{N/2}| \{ \phi \}, \{ \bar{\phi} \} \rangle
\end{align} 
Clearly for $N \rightarrow \infty$ only the largest eigenvalue dominates and so $\sigma_{L/R}$ are simply a repackaging of this maximal eigenvalue into a matrix in retarded-advance space. In our case, there are several dominant eigenvalues which seem to become degenerate in the long circumference limit. To be concrete we will assume $N$ goes to infinity last and keep only the largest of them. Conveniently, this way our results are independent of the boundary conditions. Following Ref. \cite{Schuch2012}, the entanglement spectrum is given by the spectrum of $\sqrt{\sigma^*_L} \sigma_R \sqrt{\sigma^*_L}$. The results obtained are shown in Fig. (\ref{Fig:Results}). 

\section{Relations with the AKLT wavefunction}
\label{App:AKLT}
Here we establish the relation between the AKLT wavefunction \cite{AKLT1988} and Eq. (\ref{Eq:MPS1D}), which then implies, via Eq. (\ref{Eq:Fourier}), a relation between the AKLT wavefunction and the group cohomology wavefunctions of Ref. [\onlinecite{Chen2011}]. The AKLT state is given by 
\begin{align}
\sum_{\{m\}} Tr[\Pi_i \sigma(m_i) ] |\{ m\} \rangle , 
\end{align}
with $m \in \{1,-1,0\}$ corresponding to the angular momenta of a spin-1 in the $z-$direction and $\sigma(1) = \frac{1}{\sqrt{2}} (\sigma_x + i \sigma_y)$, $\sigma(-1) = \frac{1}{\sqrt{2}} (\sigma_x - i \sigma_y)$, $\sigma(0) = \sigma_z$.  

Recall how a local unitary transformation $U^{(i)}$ acts the $i$'th site of a generic MPS  
\begin{align}
U^{(i)} &\sum_{..,g_i,..} Tr\left[... M(g) .. \right] |.. g ..\rangle  \\ \nonumber 
&=  \sum_{..,g,..} Tr\left[... M(g_i) .. \right]U^{(i)} |.. g_i ..\rangle \\ \nonumber 
&= \sum_{..,g_i g_i',..} Tr\left[... M(g_i) .. \right]U^{(i)}_{g_i'g_i} |.. g_i' ..\rangle
\end{align}
this last line can be interpreted in two ways: one can sum over $g_i'$ first, which is equivalent to rotating the basis in the $i$'th site from $|g\rangle$ to $U|g\rangle$. Alternatively one can sum over $g_i$ first meaning that the matrix associated with $g$ changes from $M(g)$ to $\tilde{M}(g') = \sum_g U_{g' g} M(g)$. Thus if one wishes to present the {\bf same} wavefunction on a {\bf different} local basis given by $U|g\rangle$, one should act with $U$ on the basis vectors and change the matrices according to 
\begin{align}
\label{Eq:MT}
\tilde{M}(g') = \sum_g U^{-1}_{g' g} M(g)
\end{align}

The minimal local unitary symmetry protecting the AKLT is $G=D_2=(1,X,Y,Z)$, consisting of $\pi$ rotations around the principal axes \cite{Pollmann2012}. As presented, the AKLT state is not written in the charge basis of this symmetry, since $|m\rangle$ transform non-trivially under the elements of $D_2$. The charge basis here is  the cartesian basis which remains invariant (up to a phase) under the action of $D_2$. The unitary transformation ($U$) which rotates a vector written in the $|-1/1/0\rangle$ basis to a vector written in the ($|x/y/z\rangle$) basis is given by 
\begin{align}
U_{ij} &= \langle x_i | m_j \rangle = \frac{1}{\sqrt{2}} \left( \begin{array}{ccc}
1 & -1 & 0 \\
i & i & 0 \\
0 & 0 & \sqrt{2} \end{array} \right) \\
|1\rangle &= \frac{1}{\sqrt{2}}(|x\rangle + i | y \rangle ) \\
|-1 \rangle &= \frac{1}{\sqrt{2}}(-|x\rangle + i | y \rangle ) \\ 
|0 \rangle &= |z\rangle 
\end{align}

According to Eq. (\ref{Eq:MT}) the matrices in the AKLT MPS transform to  
 \begin{align}
 \sigma(1) &\rightarrow  \frac{1}{\sqrt{2}}[\sigma(1) -i \sigma(-1)] = \frac{\sqrt{\bar{i}}}{\sqrt{2}}( \sigma_x -\sigma_y) \equiv \sqrt{\bar{i}} \sigma_{x-y} \\
 \sigma(-1) &\rightarrow  \frac{-1}{\sqrt{2}}[\sigma(1) +i \sigma(-1)] = \frac{-\sqrt{i}}{\sqrt{2}}( \sigma_x + \sigma_y)  \equiv -\sqrt{i} \sigma_{x+y} \\
 \sigma_z &\rightarrow \sigma_z 
 \end{align}
 with $\sqrt{i} = \frac{1+i}{\sqrt{2}}$ and $\sqrt{\bar{i}} = \frac{1-i}{\sqrt{2}}$ 

A clear difference between the our wavefunctions and the group cohomology wavefunctions compared to the AKLT wavefunction is that the dimension of the local Hilbert space is four rather than three. To interpolate between these Hilbert spaces, one should think of the spin-1 degree of freedom of the AKLT as two spin-1/2 degrees of freedom and include both the triplet and the singlet.  Accordingly, we introducing a state $|0\rangle$ associated with the identity matrix $\sigma_0$ and gradually penalized its appearance by some small coefficient $\epsilon$ (i.e. its MPS matrix is $\epsilon \sigma_0$). Gradually increasing $\epsilon$ from zero to one then performs the desired extrapolation.  By analyzing the transfer matrix of the MPS, given by $T_{\epsilon} = \epsilon^2 I\otimes I + \sigma_{x-y} \otimes \sigma_{x+y} + \sigma_{x+y} \otimes \sigma_{x-y} + \sigma_z \otimes \sigma_z$, one deduces the correlation length from the inverse of the logarithm of the ratio of the magnitudes of the largest eigenvalue over the next largest eigenvalue. For $\epsilon=0$ this ratio is $3/|-1|$, and therefore for finite $\epsilon$ it is $|3+\epsilon^2|/|-1+\epsilon^2|$. Consequently, the correlation length decreases to zero at $\epsilon=1$, which implies that nonadjacent sites are completely uncorrelated. We believe that such an extrapolation between short-range entangled and symmetry respecting wavefunctions can be lifted, via the notion of parent Hamiltonians \cite{Wolf2006, Schuch2010}, to an adiabatic path between the AKLT Hamiltonian and the extended-AKLT Hamiltonian. Proving this is however out of the scope of this appendix. 

The matrices in the extended-AKLT MPS furnish a projective representation ($\tilde{R}(g)$) of $D_2$, given by $\tilde{R}(I) = \sigma_{0}$, $\tilde{R}(X) = \sqrt{\bar{i}} \sigma_{x-y}$, $\tilde{R}(Y) = -\sqrt{i} \sigma_{x+y}$, and $\tilde{R}(Z) = \sigma_z$.  The phase factor associated with this projective representation ($\omega(g,g')$, with $g,g' \in \{I,X,Y,Z\}$) can be calculated form $\tilde{R}(g)\tilde{R}(g') = \tilde{\omega}(g,g')\tilde{R}(gg')$ yielding  
\begin{align}
&\tilde{\omega}(g \neq I,g' \neq I) = -\tilde{\omega}(g',g) \\
&\tilde{\omega}(I,g) = \tilde{\omega}(g,I) = 1 \\
&\tilde{\omega}(X,X) = -i \\
&\tilde{\omega}(Y,Y) = i \\
&\tilde{\omega}(Z,Z) = 1 \\
&\tilde{\omega}(X,Y) = -i \\
&\tilde{\omega}(X,Z) = 1 \\
&\tilde{\omega}(Y,Z) =  1
\end{align}

In the bulk, or equivalently with periodic boundary conditions, the size of the matrices making up the MPS is irrelevant and only their algebra affects the state. As shown in Eq. (\ref{Eq:RRep}), the matrices ($R(g)$) of the MPS in Eq. (\ref{Eq:MPS1D}), also yield a projective representation of $Z_2 \times Z_2 \simeq D_2$. Two projective representations are called equivalent if they can be obtained from one another by $R(g) \simeq \Phi(g) \tilde{R}(g)$, where $\Phi(g)$ is a function from $g \in D_2$ to $U(1)$ and the equality is at the level of the algebra. In the MPS context, this equivalence reflects the gauge freedom in choosing the phases of our basis vector $|g\rangle$. Two equivalent projective representations would thus generate the same state, up to a local symmetry respecting gauge transformation. 

The remaining task is thus to show that a $\Phi(g)$ connecting the two representations exists. First we fix our association of elements in $[\sigma,\tau] \in Z_2 \times Z_2$ with $g \in D_2$ to be $[0,0] = I,[1,0]=X,[0,1]=Y$ and $[1,1]=Z$. Following Eq. (\ref{Eq:RRep}) with $m=1$, the phase factor $\omega$ is given by 
\begin{align}
&\omega(g \neq I,g' \neq I) = -\omega(g',g) \\
&\omega(I,g) = \omega(g,I) = 1 \\
&\omega(X,X) = 1 \\
&\omega(Y,Y) = 1 \\
&\omega(Z,Z) = -1 \\
&\omega(X,Y) = 1 \\
&\omega(X,Z) = 1 \\
&\omega(Y,Z) =  -1
\end{align} 
As one can verify that using $\Phi(I) = 1$, $\Phi(X) = \frac{1+i}{\sqrt{2}} = \sqrt{i}$, $\Phi(Y) = -\sqrt{\bar{i}}$, and $\Phi(Z) = i$, one obtains that $\omega(g,g') = \tilde{\omega}(g,g') \Phi(g) \Phi(g')/\Phi(gg')$, and thus $R(g) \simeq \Phi(g) \tilde{R}(g)$ as required. 

\section{Deriving the TPS ansatz from a $U(1)$ SPT wavefunction} 
\label{App:FromSPT}
The TPS ansatz in Eq. (\ref{Eq:TPS}) maybe also be derived starting from the bosonic 2D $U(1)$ SPT wavefunction of Ref. [\onlinecite{Senthil2013}] 
\begin{align}
\Psi_{mod} &= \Pi_{i<j} |z_i - z_j| |w_i-w_j|  \Pi_{i,j} \frac{(z_i - w_j)}{|z_i-w_j|}e^{-\sum_i \frac{|z_i|^2+|w_i|^2}{4}}
\end{align}
Analogously to what was done for the Laughlin wavefunction in App. (\ref{App:LaughlinTPS}), $\psi_{mod}$ can be presented as a continuum TPS using two virtual $U(1)$ fields ($\phi,\varphi \in [0..1)$) associated with the two types of particles. This results in the following expression 
\begin{align}
\Psi_{mod}(\{ z_i \},\{ w_j \}) &= \int{\rm D} \phi e^{ -S[\phi]-S[\varphi]} \Pi_i A_i B_i \\ \nonumber 
A_i &= \delta \left(\int_{O_{z_i}} dl \nabla \phi - 1 \right), \\ \nonumber
B_i &= \delta \left(\int_{O_{w_i}} dl \nabla \varphi - 1 \right), \\ \nonumber 
S[\phi] &= \int d^2 r \frac{\pi}{2} (\nabla \phi)^2 + 2\pi i \phi [\rho_w(r)-\rho_{0}], \\ \nonumber 
S[\varphi] &= \int d^2 r \frac{\pi}{2} (\nabla \varphi)^2 + 2\pi i \varphi [\rho_z(r)-\rho_0]
\end{align}
where $\rho_w(r) = \sum_i \delta(r-w_i)$, $\rho_z(r) = \sum_i \delta(r-z_i)$, and $\rho_0$ is the average density of the $z_i$'s or the $w_i$'s charges. 

The wavefunction $\Psi_{mod}$ actually obeys a $U(1) \times U(1)$ symmetry, since it conserves both types of charges separately. The $U(1)$ symmetry that preserves the SPT is associated with the charge difference while the symmetry associated with the total charge is superfluous here. To disentangle these two parts, we can rewrite the above continuum TPS using the fields $\Theta = \phi + \varphi$ and $\theta = \phi-\varphi$, both being periodic in the interval $[0..1)$. This results in 
\begin{align}
\Psi_{mod}((\{ z_i \},\{ w_j \}) &= \int{\rm D} \phi e^{ -S[\Theta]-S[\theta]} \Pi_i C_i c_i \\ \nonumber 
C_i &= \delta \left(\int_{O_{z_i}} dl \nabla \Theta - 1 \right)\delta \left(\int_{O_{w_i}} dl \nabla \Theta - 1 \right) \\ \nonumber 
c_i &= \delta \left(\int_{O_{z_i}} dl \nabla \theta- 1 \right)\delta \left(\int_{O_{w_i}} dl \nabla \theta + 1 \right) \\ \nonumber
S[\Theta] &= \int d^2 r \frac{\pi}{4} (\nabla \Theta)^2 + i \pi \Theta (\Sigma(r)-2 \rho_{0}), \\ \nonumber 
S[\theta] &= \int d^2 r \frac{\pi}{4} (\nabla \theta)^2 + i \pi \theta \rho(r)
\end{align}
where $\Sigma(r) = \rho_z(r) + \rho_w(r)$ and $\rho(r) = \rho_z - \rho_w$. 

Next we break the superfluous $U(1)$ symmetry associated with the total charge, by making a condensate of $(z,w)$ pairs. Previously $\Sigma(r)$ and $\rho(r)$ were dependent by the restriction that the original particle density cannot be negative or equivalently $|\int_{region} \rho(r)| \leq |\int_{region} \Sigma(r)|$. As this pair-condensate density increases, this constrain becomes less significant and we thus consider $\Sigma(r)$ and $\rho(r)$ as independent quantities. The wavefunction then factorizes into a product of $\Psi_{superfluous}$ and $\Psi_{SPT}$, the latter given by 
\begin{align}
\Psi_{SPT}(\rho) &= \int{\rm D} \phi e^{-S[\theta]} \Pi_i c_i \\ \nonumber 
c_i &= \delta \left(\int_{O_{x^+_i}} dl \nabla \theta- 1 \right)\delta \left(\int_{O_{x^{-}_i}} dl \nabla \theta + 1 \right) \\ \nonumber
S[\theta] &= \int d^2 r \frac{\pi}{4} (\nabla \theta)^2 + i  \pi \theta \rho(r)
\end{align}
where $\rho(r) = \sum_{i} \delta(r-x^{+}_i) - \delta(r-x^{-}_i)$. The above continuum TPS can now be discretized by taking $U(1) \rightarrow Z_N$ and space to a lattice similarly to what has been done in App. (\ref{App:LaughlinTPS}). A pleasant feature of this $U(1)$-SPT viewpoint is that the divergent circulation of the $U(1)$ vorticity is avoided since $\int d^2 r \rho(r) = 0$. Consequently we see no obvious obstruction to placing this continuum TPS, with the full $U(1)$ symmetry, on a lattice. This we could not do for the continuum TPS of the Laughlin state in Eq. (\ref{AppEq:TPS1Q})

\bibliography{SPTRefs}

\begin{thebibliography}{48}
\expandafter\ifx\csname natexlab\endcsname\relax\def\natexlab#1{#1}\fi
\expandafter\ifx\csname bibnamefont\endcsname\relax
  \def\bibnamefont#1{#1}\fi
\expandafter\ifx\csname bibfnamefont\endcsname\relax
  \def\bibfnamefont#1{#1}\fi
\expandafter\ifx\csname citenamefont\endcsname\relax
  \def\citenamefont#1{#1}\fi
\expandafter\ifx\csname url\endcsname\relax
  \def\url#1{\texttt{#1}}\fi
\expandafter\ifx\csname urlprefix\endcsname\relax\def\urlprefix{URL }\fi
\providecommand{\bibinfo}[2]{#2}
\providecommand{\eprint}[2][]{\url{#2}}

\bibitem[{\citenamefont{Chen et~al.}(2013)\citenamefont{Chen, Gu, Liu, and
  Wen}}]{Chen2011}
\bibinfo{author}{\bibfnamefont{X.}~\bibnamefont{Chen}},
  \bibinfo{author}{\bibfnamefont{Z.-C.} \bibnamefont{Gu}},
  \bibinfo{author}{\bibfnamefont{Z.-X.} \bibnamefont{Liu}}, \bibnamefont{and}
  \bibinfo{author}{\bibfnamefont{X.-G.} \bibnamefont{Wen}},
  \bibinfo{journal}{Phys. Rev. B} \textbf{\bibinfo{volume}{87}},
  \bibinfo{pages}{155114} (\bibinfo{year}{2013}),
  \urlprefix\url{http://link.aps.org/doi/10.1103/PhysRevB.87.155114}.

\bibitem[{\citenamefont{Levin and Gu}(2012)}]{Levin2012}
\bibinfo{author}{\bibfnamefont{M.}~\bibnamefont{Levin}} \bibnamefont{and}
  \bibinfo{author}{\bibfnamefont{Z.-C.} \bibnamefont{Gu}},
  \bibinfo{journal}{Phys. Rev. B} \textbf{\bibinfo{volume}{86}},
  \bibinfo{pages}{115109} (\bibinfo{year}{2012}),
  \urlprefix\url{http://link.aps.org/doi/10.1103/PhysRevB.86.115109}.

\bibitem[{\citenamefont{Lu and Vishwanath}(2012)}]{YuanMing2012}
\bibinfo{author}{\bibfnamefont{Y.-M.} \bibnamefont{Lu}} \bibnamefont{and}
  \bibinfo{author}{\bibfnamefont{A.}~\bibnamefont{Vishwanath}},
  \bibinfo{journal}{Phys. Rev. B} \textbf{\bibinfo{volume}{86}},
  \bibinfo{pages}{125119} (\bibinfo{year}{2012}),
  \urlprefix\url{http://link.aps.org/doi/10.1103/PhysRevB.86.125119}.

\bibitem[{\citenamefont{{Kapustin}}(2014)}]{Kapustin2014}
\bibinfo{author}{\bibfnamefont{A.}~\bibnamefont{{Kapustin}}},
  \bibinfo{journal}{ArXiv e-prints}  (\bibinfo{year}{2014}),
  \eprint{1403.1467}.

\bibitem[{\citenamefont{Schuch et~al.}(2011)\citenamefont{Schuch,
  P\'erez-Garc\'~ia, and Cirac}}]{Schuch2011}
\bibinfo{author}{\bibfnamefont{N.}~\bibnamefont{Schuch}},
  \bibinfo{author}{\bibfnamefont{D.}~\bibnamefont{P\'erez-Garc\'~ia}},
  \bibnamefont{and} \bibinfo{author}{\bibfnamefont{I.}~\bibnamefont{Cirac}},
  \bibinfo{journal}{Phys. Rev. B} \textbf{\bibinfo{volume}{84}},
  \bibinfo{pages}{165139} (\bibinfo{year}{2011}),
  \urlprefix\url{http://link.aps.org/doi/10.1103/PhysRevB.84.165139}.

\bibitem[{\citenamefont{Hasan and Kane}(2010)}]{Hasan2010}
\bibinfo{author}{\bibfnamefont{M.~Z.} \bibnamefont{Hasan}} \bibnamefont{and}
  \bibinfo{author}{\bibfnamefont{C.~L.} \bibnamefont{Kane}},
  \bibinfo{journal}{Reviews of Modern Physics} \textbf{\bibinfo{volume}{82}},
  \bibinfo{pages}{3045} (\bibinfo{year}{2010}).

\bibitem[{\citenamefont{Buyers et~al.}(1986)\citenamefont{Buyers, Morra,
  Armstrong, Hogan, Gerlach, and Hirakawa}}]{Buyers1986}
\bibinfo{author}{\bibfnamefont{W.~J.~L.} \bibnamefont{Buyers}},
  \bibinfo{author}{\bibfnamefont{R.~M.} \bibnamefont{Morra}},
  \bibinfo{author}{\bibfnamefont{R.~L.} \bibnamefont{Armstrong}},
  \bibinfo{author}{\bibfnamefont{M.~J.} \bibnamefont{Hogan}},
  \bibinfo{author}{\bibfnamefont{P.}~\bibnamefont{Gerlach}}, \bibnamefont{and}
  \bibinfo{author}{\bibfnamefont{K.}~\bibnamefont{Hirakawa}},
  \bibinfo{journal}{Phys. Rev. Lett.} \textbf{\bibinfo{volume}{56}},
  \bibinfo{pages}{371} (\bibinfo{year}{1986}),
  \urlprefix\url{http://link.aps.org/doi/10.1103/PhysRevLett.56.371}.

\bibitem[{\citenamefont{{Liu} et~al.}(2014)\citenamefont{{Liu}, {Gu}, and
  {Wen}}}]{Wen2014}
\bibinfo{author}{\bibfnamefont{Z.-X.} \bibnamefont{{Liu}}},
  \bibinfo{author}{\bibfnamefont{Z.-C.} \bibnamefont{{Gu}}}, \bibnamefont{and}
  \bibinfo{author}{\bibfnamefont{X.-G.} \bibnamefont{{Wen}}},
  \bibinfo{journal}{ArXiv e-prints}  (\bibinfo{year}{2014}),
  \eprint{1404.2818}.

\bibitem[{\citenamefont{Halperin}(1983)}]{Halperin1983}
\bibinfo{author}{\bibfnamefont{B.~I.} \bibnamefont{Halperin}},
  \bibinfo{journal}{Helvetica Physica Acta} \textbf{\bibinfo{volume}{56}},
  \bibinfo{pages}{75} (\bibinfo{year}{1983}).

\bibitem[{\citenamefont{Senthil and Levin}(2013)}]{Senthil2013}
\bibinfo{author}{\bibfnamefont{T.}~\bibnamefont{Senthil}} \bibnamefont{and}
  \bibinfo{author}{\bibfnamefont{M.}~\bibnamefont{Levin}},
  \bibinfo{journal}{Phys. Rev. Lett.} \textbf{\bibinfo{volume}{110}},
  \bibinfo{pages}{046801} (\bibinfo{year}{2013}),
  \urlprefix\url{http://link.aps.org/doi/10.1103/PhysRevLett.110.046801}.

\bibitem[{\citenamefont{{Lu} and {Vishwanath}}(2013)}]{YuanMing2013}
\bibinfo{author}{\bibfnamefont{Y.-M.} \bibnamefont{{Lu}}} \bibnamefont{and}
  \bibinfo{author}{\bibfnamefont{A.}~\bibnamefont{{Vishwanath}}},
  \bibinfo{journal}{ArXiv e-prints}  (\bibinfo{year}{2013}),
  \eprint{1302.2634}.

\bibitem[{\citenamefont{Chen et~al.}(2014)\citenamefont{Chen, Lu, and
  Vishwanath}}]{AshvinDecorated}
\bibinfo{author}{\bibfnamefont{X.}~\bibnamefont{Chen}},
  \bibinfo{author}{\bibfnamefont{Y.-M.} \bibnamefont{Lu}}, \bibnamefont{and}
  \bibinfo{author}{\bibfnamefont{A.}~\bibnamefont{Vishwanath}},
  \bibinfo{journal}{Nat Commun} \textbf{\bibinfo{volume}{5}}
  (\bibinfo{year}{2014}), \urlprefix\url{http://dx.doi.org/10.1038/ncomms4507}.

\bibitem[{\citenamefont{{Zaletel} et~al.}(2014)\citenamefont{{Zaletel}, {Mong},
  and {Pollmann}}}]{Zaletel2014}
\bibinfo{author}{\bibfnamefont{M.~P.} \bibnamefont{{Zaletel}}},
  \bibinfo{author}{\bibfnamefont{R.~S.~K.} \bibnamefont{{Mong}}},
  \bibnamefont{and}
  \bibinfo{author}{\bibfnamefont{F.}~\bibnamefont{{Pollmann}}},
  \bibinfo{journal}{ArXiv e-prints}  (\bibinfo{year}{2014}),
  \eprint{1405.6028}.

\bibitem[{\citenamefont{Santos and Wang}(2014)}]{Juven1}
\bibinfo{author}{\bibfnamefont{L.~H.} \bibnamefont{Santos}} \bibnamefont{and}
  \bibinfo{author}{\bibfnamefont{J.}~\bibnamefont{Wang}},
  \bibinfo{journal}{Phys. Rev. B} \textbf{\bibinfo{volume}{89}},
  \bibinfo{pages}{195122} (\bibinfo{year}{2014}),
  \urlprefix\url{http://link.aps.org/doi/10.1103/PhysRevB.89.195122}.

\bibitem[{\citenamefont{{Wang} et~al.}(2014)\citenamefont{{Wang}, {Santos}, and
  {Wen}}}]{Juven2}
\bibinfo{author}{\bibfnamefont{J.}~\bibnamefont{{Wang}}},
  \bibinfo{author}{\bibfnamefont{L.~H.} \bibnamefont{{Santos}}},
  \bibnamefont{and} \bibinfo{author}{\bibfnamefont{X.-G.} \bibnamefont{{Wen}}},
  \bibinfo{journal}{ArXiv e-prints}  (\bibinfo{year}{2014}),
  \eprint{1403.5256}.

\bibitem[{\citenamefont{Laughlin}(1983)}]{Laughlin1983}
\bibinfo{author}{\bibfnamefont{R.~B.} \bibnamefont{Laughlin}},
  \bibinfo{journal}{Phys. Rev. Lett.} \textbf{\bibinfo{volume}{50}},
  \bibinfo{pages}{1395} (\bibinfo{year}{1983}),
  \urlprefix\url{http://link.aps.org/doi/10.1103/PhysRevLett.50.1395}.

\bibitem[{\citenamefont{Zhang}(1992)}]{Zhang1992}
\bibinfo{author}{\bibfnamefont{S.~C.} \bibnamefont{Zhang}},
  \bibinfo{journal}{International Journal of Modern Physics B}
  \textbf{\bibinfo{volume}{06}}, \bibinfo{pages}{803} (\bibinfo{year}{1992}).

\bibitem[{\citenamefont{Jain}(2007)}]{Jain2007}
\bibinfo{author}{\bibfnamefont{J.}~\bibnamefont{Jain}},
  \emph{\bibinfo{title}{Composite Fermions}} (\bibinfo{publisher}{Cambridge
  University Press}, \bibinfo{year}{2007}), ISBN \bibinfo{isbn}{9781139462648},
  \urlprefix\url{http://books.google.co.il/books?id=0jv9UF6UL20C}.

\bibitem[{\citenamefont{Haldane}(1983)}]{Haldane1983}
\bibinfo{author}{\bibfnamefont{F.~D.~M.} \bibnamefont{Haldane}},
  \bibinfo{journal}{Phys. Rev. Lett.} \textbf{\bibinfo{volume}{51}},
  \bibinfo{pages}{605} (\bibinfo{year}{1983}),
  \urlprefix\url{http://link.aps.org/doi/10.1103/PhysRevLett.51.605}.

\bibitem[{\citenamefont{Trugman and Kivelson}(1985)}]{Trugman1985}
\bibinfo{author}{\bibfnamefont{S.~A.} \bibnamefont{Trugman}} \bibnamefont{and}
  \bibinfo{author}{\bibfnamefont{S.}~\bibnamefont{Kivelson}},
  \bibinfo{journal}{Phys. Rev. B} \textbf{\bibinfo{volume}{31}},
  \bibinfo{pages}{5280} (\bibinfo{year}{1985}),
  \urlprefix\url{http://link.aps.org/doi/10.1103/PhysRevB.31.5280}.

\bibitem[{\citenamefont{Chakraborty and
  Pietil{\"a}inen}(1988)}]{Chakraborty1988}
\bibinfo{author}{\bibfnamefont{T.}~\bibnamefont{Chakraborty}} \bibnamefont{and}
  \bibinfo{author}{\bibfnamefont{P.}~\bibnamefont{Pietil{\"a}inen}},
  \emph{\bibinfo{title}{The fractional quantum Hall effect: properties of an
  incompressible quantum fluid}}, Springer series in solid-state sciences
  (\bibinfo{publisher}{Springer-Verlag}, \bibinfo{year}{1988}), ISBN
  \bibinfo{isbn}{9783540191117},
  \urlprefix\url{http://books.google.co.uk/books?id=oY4fAQAAMAAJ}.

\bibitem[{\citenamefont{Zaletel and Mong}(2012)}]{Zaletel2012}
\bibinfo{author}{\bibfnamefont{M.~P.} \bibnamefont{Zaletel}} \bibnamefont{and}
  \bibinfo{author}{\bibfnamefont{R.~S.~K.} \bibnamefont{Mong}},
  \bibinfo{journal}{Phys. Rev. B} \textbf{\bibinfo{volume}{86}},
  \bibinfo{pages}{245305} (\bibinfo{year}{2012}),
  \urlprefix\url{http://link.aps.org/doi/10.1103/PhysRevB.86.245305}.

\bibitem[{\citenamefont{Moore and Read}(1991)}]{Moore1991}
\bibinfo{author}{\bibfnamefont{G.}~\bibnamefont{Moore}} \bibnamefont{and}
  \bibinfo{author}{\bibfnamefont{N.}~\bibnamefont{Read}},
  \bibinfo{journal}{Nuclear Physics B} \textbf{\bibinfo{volume}{360}},
  \bibinfo{pages}{362 } (\bibinfo{year}{1991}), ISSN \bibinfo{issn}{0550-3213},
  \urlprefix\url{http://www.sciencedirect.com/science/article/pii/055032139190407O}.

\bibitem[{\citenamefont{Girvin and MacDonald}(1987)}]{Girvin1987}
\bibinfo{author}{\bibfnamefont{S.~M.} \bibnamefont{Girvin}} \bibnamefont{and}
  \bibinfo{author}{\bibfnamefont{A.~H.} \bibnamefont{MacDonald}},
  \bibinfo{journal}{Phys. Rev. Lett.} \textbf{\bibinfo{volume}{58}},
  \bibinfo{pages}{1252} (\bibinfo{year}{1987}),
  \urlprefix\url{http://link.aps.org/doi/10.1103/PhysRevLett.58.1252}.

\bibitem[{\citenamefont{Read}(1989)}]{Read1989}
\bibinfo{author}{\bibfnamefont{N.}~\bibnamefont{Read}}, \bibinfo{journal}{Phys.
  Rev. Lett.} \textbf{\bibinfo{volume}{62}}, \bibinfo{pages}{86}
  (\bibinfo{year}{1989}),
  \urlprefix\url{http://link.aps.org/doi/10.1103/PhysRevLett.62.86}.

\bibitem[{\citenamefont{Kennedy and Tasaki}(1992)}]{Kennedy1992}
\bibinfo{author}{\bibfnamefont{T.}~\bibnamefont{Kennedy}} \bibnamefont{and}
  \bibinfo{author}{\bibfnamefont{H.}~\bibnamefont{Tasaki}},
  \bibinfo{journal}{Communications in Mathematical Physics}
  \textbf{\bibinfo{volume}{147}}, \bibinfo{pages}{431} (\bibinfo{year}{1992}),
  ISSN \bibinfo{issn}{0010-3616},
  \urlprefix\url{http://dx.doi.org/10.1007/BF02097239}.

\bibitem[{\citenamefont{Else et~al.}(2013)\citenamefont{Else, Bartlett, and
  Doherty}}]{Else2013}
\bibinfo{author}{\bibfnamefont{D.~V.} \bibnamefont{Else}},
  \bibinfo{author}{\bibfnamefont{S.~D.} \bibnamefont{Bartlett}},
  \bibnamefont{and} \bibinfo{author}{\bibfnamefont{A.~C.}
  \bibnamefont{Doherty}}, \bibinfo{journal}{Phys. Rev. B}
  \textbf{\bibinfo{volume}{88}}, \bibinfo{pages}{085114}
  (\bibinfo{year}{2013}),
  \urlprefix\url{http://link.aps.org/doi/10.1103/PhysRevB.88.085114}.

\bibitem[{\citenamefont{Duivenvoorden and Quella}(2013)}]{Quella2013}
\bibinfo{author}{\bibfnamefont{K.}~\bibnamefont{Duivenvoorden}}
  \bibnamefont{and} \bibinfo{author}{\bibfnamefont{T.}~\bibnamefont{Quella}},
  \bibinfo{journal}{Phys. Rev. B} \textbf{\bibinfo{volume}{88}},
  \bibinfo{pages}{125115} (\bibinfo{year}{2013}),
  \urlprefix\url{http://link.aps.org/doi/10.1103/PhysRevB.88.125115}.

\bibitem[{\citenamefont{Chen et~al.}(2011)\citenamefont{Chen, Liu, and
  Wen}}]{Chen2011b}
\bibinfo{author}{\bibfnamefont{X.}~\bibnamefont{Chen}},
  \bibinfo{author}{\bibfnamefont{Z.-X.} \bibnamefont{Liu}}, \bibnamefont{and}
  \bibinfo{author}{\bibfnamefont{X.-G.} \bibnamefont{Wen}},
  \bibinfo{journal}{Phys. Rev. B} \textbf{\bibinfo{volume}{84}},
  \bibinfo{pages}{235141} (\bibinfo{year}{2011}),
  \urlprefix\url{http://link.aps.org/doi/10.1103/PhysRevB.84.235141}.

\bibitem[{\citenamefont{{Hastings}}(2014)}]{Hastings2014}
\bibinfo{author}{\bibfnamefont{M.~B.} \bibnamefont{{Hastings}}},
  \bibinfo{journal}{ArXiv e-prints}  (\bibinfo{year}{2014}),
  \eprint{1404.4327}.

\bibitem[{tra()}]{transgression}
\bibinfo{note}{This map between projective representations and regular ones is
  called transgression in the mathematical literature \cite{Gerbs}}.

\bibitem[{\citenamefont{{Schuch} et~al.}(2012)\citenamefont{{Schuch},
  {Poilblanc}, {Cirac}, and {Perez-Garcia}}}]{Schuch2012}
\bibinfo{author}{\bibfnamefont{N.}~\bibnamefont{{Schuch}}},
  \bibinfo{author}{\bibfnamefont{D.}~\bibnamefont{{Poilblanc}}},
  \bibinfo{author}{\bibfnamefont{J.~I.} \bibnamefont{{Cirac}}},
  \bibnamefont{and}
  \bibinfo{author}{\bibfnamefont{D.}~\bibnamefont{{Perez-Garcia}}},
  \bibinfo{journal}{ArXiv e-prints}  (\bibinfo{year}{2012}),
  \eprint{1210.5601}.

\bibitem[{\citenamefont{Hatcher}(2002)}]{Hatcher2002}
\bibinfo{author}{\bibfnamefont{A.}~\bibnamefont{Hatcher}},
  \emph{\bibinfo{title}{Algebraic Topology}} (\bibinfo{publisher}{Cambridge
  University Press}, \bibinfo{year}{2002}), ISBN \bibinfo{isbn}{9780521795401},
  \urlprefix\url{http://books.google.it/books?id=BjKs86kosqgC}.

\bibitem[{\citenamefont{{Kapustin} and {Thorngren}}(2014)}]{Ryan2014}
\bibinfo{author}{\bibfnamefont{A.}~\bibnamefont{{Kapustin}}} \bibnamefont{and}
  \bibinfo{author}{\bibfnamefont{R.}~\bibnamefont{{Thorngren}}},
  \bibinfo{journal}{ArXiv e-prints}  (\bibinfo{year}{2014}),
  \eprint{1404.3230}.

\bibitem[{\citenamefont{{Verstraete} and {Cirac}}(2004)}]{Verstraete2004}
\bibinfo{author}{\bibfnamefont{F.}~\bibnamefont{{Verstraete}}}
  \bibnamefont{and} \bibinfo{author}{\bibfnamefont{J.~I.}
  \bibnamefont{{Cirac}}}, \bibinfo{journal}{eprint arXiv:cond-mat/0407066}
  (\bibinfo{year}{2004}), \eprint{cond-mat/0407066}.

\bibitem[{\citenamefont{{Perez-Garcia}
  et~al.}(2007)\citenamefont{{Perez-Garcia}, {Verstraete}, {Cirac}, and
  {Wolf}}}]{Perez-Garcia2007}
\bibinfo{author}{\bibfnamefont{D.}~\bibnamefont{{Perez-Garcia}}},
  \bibinfo{author}{\bibfnamefont{F.}~\bibnamefont{{Verstraete}}},
  \bibinfo{author}{\bibfnamefont{J.~I.} \bibnamefont{{Cirac}}},
  \bibnamefont{and} \bibinfo{author}{\bibfnamefont{M.~M.}
  \bibnamefont{{Wolf}}}, \bibinfo{journal}{ArXiv e-prints}
  (\bibinfo{year}{2007}), \eprint{0707.2260}.

\bibitem[{\citenamefont{You et~al.}(2014)\citenamefont{You, Bi, Rasmussen,
  Slagle, and Xu}}]{Cenke2014}
\bibinfo{author}{\bibfnamefont{Y.-Z.} \bibnamefont{You}},
  \bibinfo{author}{\bibfnamefont{Z.}~\bibnamefont{Bi}},
  \bibinfo{author}{\bibfnamefont{A.}~\bibnamefont{Rasmussen}},
  \bibinfo{author}{\bibfnamefont{K.}~\bibnamefont{Slagle}}, \bibnamefont{and}
  \bibinfo{author}{\bibfnamefont{C.}~\bibnamefont{Xu}}, \bibinfo{journal}{Phys.
  Rev. Lett.} \textbf{\bibinfo{volume}{112}}, \bibinfo{pages}{247202}
  (\bibinfo{year}{2014}),
  \urlprefix\url{http://link.aps.org/doi/10.1103/PhysRevLett.112.247202}.

\bibitem[{\citenamefont{Li and Haldane}(2008)}]{Li2008}
\bibinfo{author}{\bibfnamefont{H.}~\bibnamefont{Li}} \bibnamefont{and}
  \bibinfo{author}{\bibfnamefont{F.~D.~M.} \bibnamefont{Haldane}},
  \bibinfo{journal}{Phys. Rev. Lett.} \textbf{\bibinfo{volume}{101}},
  \bibinfo{pages}{010504} (\bibinfo{year}{2008}),
  \urlprefix\url{http://link.aps.org/doi/10.1103/PhysRevLett.101.010504}.

\bibitem[{\citenamefont{Fidkowski}(2010)}]{Fidkowski2010}
\bibinfo{author}{\bibfnamefont{L.}~\bibnamefont{Fidkowski}},
  \bibinfo{journal}{Phys. Rev. Lett.} \textbf{\bibinfo{volume}{104}},
  \bibinfo{pages}{130502} (\bibinfo{year}{2010}),
  \urlprefix\url{http://link.aps.org/doi/10.1103/PhysRevLett.104.130502}.

\bibitem[{\citenamefont{Kapit and Mueller}(2010)}]{Kapit2010}
\bibinfo{author}{\bibfnamefont{E.}~\bibnamefont{Kapit}} \bibnamefont{and}
  \bibinfo{author}{\bibfnamefont{E.}~\bibnamefont{Mueller}},
  \bibinfo{journal}{Phys. Rev. Lett.} \textbf{\bibinfo{volume}{105}},
  \bibinfo{pages}{215303} (\bibinfo{year}{2010}),
  \urlprefix\url{http://link.aps.org/doi/10.1103/PhysRevLett.105.215303}.

\bibitem[{\citenamefont{Scaffidi and Simon}(2014)}]{Scaffidi2014}
\bibinfo{author}{\bibfnamefont{T.}~\bibnamefont{Scaffidi}} \bibnamefont{and}
  \bibinfo{author}{\bibfnamefont{S.~H.} \bibnamefont{Simon}},
  \bibinfo{journal}{Phys. Rev. B} \textbf{\bibinfo{volume}{90}},
  \bibinfo{pages}{115132} (\bibinfo{year}{2014}),
  \urlprefix\url{http://link.aps.org/doi/10.1103/PhysRevB.90.115132}.

\bibitem[{\citenamefont{Kosterlitz and Thouless}(1973)}]{Kosterlitz1973}
\bibinfo{author}{\bibfnamefont{J.~M.} \bibnamefont{Kosterlitz}}
  \bibnamefont{and} \bibinfo{author}{\bibfnamefont{D.~J.}
  \bibnamefont{Thouless}}, \bibinfo{journal}{Journal of Physics C: Solid State
  Physics} \textbf{\bibinfo{volume}{6}}, \bibinfo{pages}{1181}
  (\bibinfo{year}{1973}),
  \urlprefix\url{http://stacks.iop.org/0022-3719/6/i=7/a=010}.

\bibitem[{\citenamefont{Polyakov}(1987)}]{Polyakov1987}
\bibinfo{author}{\bibfnamefont{A.~M.} \bibnamefont{Polyakov}},
  \emph{\bibinfo{title}{Gauge Fields and Strings}}, Contemporary concepts in
  physics (\bibinfo{publisher}{Taylor \& Francis}, \bibinfo{year}{1987}), ISBN
  \bibinfo{isbn}{9783718603930}.

\bibitem[{\citenamefont{Affleck et~al.}(1988)\citenamefont{Affleck, Kennedy,
  Lieb, and Tasaki}}]{AKLT1988}
\bibinfo{author}{\bibfnamefont{I.}~\bibnamefont{Affleck}},
  \bibinfo{author}{\bibfnamefont{T.}~\bibnamefont{Kennedy}},
  \bibinfo{author}{\bibfnamefont{E.~H.} \bibnamefont{Lieb}}, \bibnamefont{and}
  \bibinfo{author}{\bibfnamefont{H.}~\bibnamefont{Tasaki}},
  \bibinfo{journal}{Comm. Math. Phys.} \textbf{\bibinfo{volume}{115}},
  \bibinfo{pages}{477} (\bibinfo{year}{1988}),
  \urlprefix\url{http://projecteuclid.org/euclid.cmp/1104161001}.

\bibitem[{\citenamefont{Pollmann et~al.}(2012)\citenamefont{Pollmann, Berg,
  Turner, and Oshikawa}}]{Pollmann2012}
\bibinfo{author}{\bibfnamefont{F.}~\bibnamefont{Pollmann}},
  \bibinfo{author}{\bibfnamefont{E.}~\bibnamefont{Berg}},
  \bibinfo{author}{\bibfnamefont{A.~M.} \bibnamefont{Turner}},
  \bibnamefont{and} \bibinfo{author}{\bibfnamefont{M.}~\bibnamefont{Oshikawa}},
  \bibinfo{journal}{Phys. Rev. B} \textbf{\bibinfo{volume}{85}},
  \bibinfo{pages}{075125} (\bibinfo{year}{2012}),
  \urlprefix\url{http://link.aps.org/doi/10.1103/PhysRevB.85.075125}.

\bibitem[{\citenamefont{Wolf et~al.}(2006)\citenamefont{Wolf, Ortiz,
  Verstraete, and Cirac}}]{Wolf2006}
\bibinfo{author}{\bibfnamefont{M.~M.} \bibnamefont{Wolf}},
  \bibinfo{author}{\bibfnamefont{G.}~\bibnamefont{Ortiz}},
  \bibinfo{author}{\bibfnamefont{F.}~\bibnamefont{Verstraete}},
  \bibnamefont{and} \bibinfo{author}{\bibfnamefont{J.~I.} \bibnamefont{Cirac}},
  \bibinfo{journal}{Phys. Rev. Lett.} \textbf{\bibinfo{volume}{97}},
  \bibinfo{pages}{110403} (\bibinfo{year}{2006}),
  \urlprefix\url{http://link.aps.org/doi/10.1103/PhysRevLett.97.110403}.

\bibitem[{Sch(2010)}]{Schuch2010}
\bibinfo{journal}{Annals of Physics} \textbf{\bibinfo{volume}{325}},
  \bibinfo{pages}{2153 } (\bibinfo{year}{2010}), ISSN
  \bibinfo{issn}{0003-4916},
  \urlprefix\url{http://www.sciencedirect.com/science/article/pii/S0003491610000990}.

\bibitem[{\citenamefont{{Willerton}}(2005)}]{Gerbs}
\bibinfo{author}{\bibfnamefont{S.}~\bibnamefont{{Willerton}}},
  \bibinfo{journal}{ArXiv Mathematics e-prints}  (\bibinfo{year}{2005}),
  \eprint{math/0503266}.

\end{thebibliography}
\end{document}